\documentclass[twocolumn,tighten]{aastex62}
\usepackage{amsmath}

\newcommand\be{\begin{equation}}
\newcommand\en{\end{equation}}

\newcommand\etal{{\rm et al}.\ }

\begin{document}

\title{Asymmetric Temperature Variations In Protoplanetary disks: I. Linear Theory, Corotating Spirals, and Ring Formation}

\shorttitle{Asymmetric Temperature Variations: I}
\shortauthors{Zhu et al.}

\correspondingauthor{Zhaohuan Zhu}
\email{zhaohuan.zhu@unlv.edu}

\author[0000-0003-3616-6822]{Zhaohuan Zhu}
\affiliation{Department of Physics and Astronomy, University of Nevada, Las Vegas, 4505 S.~Maryland Pkwy, Las Vegas, NV, 89154, USA}
\affiliation{Nevada Center for Astrophysics, University of Nevada, Las Vegas, 4505 S.~Maryland Pkwy, Las Vegas, NV, 89154, USA}
\author[0000-0002-8537-9114]{Shangjia Zhang}
\altaffiliation{NASA Hubble Fellowship Program (NHFP) Sagan Fellow}
\affiliation{Department of Astronomy, Columbia University, 538 W. 120th Street, Pupin Hall, New York, NY, 10027, USA}
\author[0000-0002-1570-2203]{Ted Johnson}
\affiliation{Department of Physics and Astronomy, University of Nevada, Las Vegas, 4505 S.~Maryland Pkwy, Las Vegas, NV, 89154, USA}
\affiliation{Nevada Center for Astrophysics, University of Nevada, Las Vegas, 4505 S.~Maryland Pkwy, Las Vegas, NV, 89154, USA}

\begin{abstract}

Protoplanetary disks can exhibit asymmetric temperature variations due to phenomena such as shadows cast by the inner disk or localized heating by young planets. 
We investigate the disk features induced by these asymmetric temperature variations. We find that spirals are initially excited, then break into two and reconnect to form rings. By carrying out linear analyses, we first study the spiral launching mechanism, and find that the effects of azimuthal temperature variations share similarities with effects of external potentials. Specifically, rotating temperature variations launch steady spiral structures at Lindblad resonances, which corotate with the temperature patterns.  When the cooling time exceeds the orbital period, these spiral structures are significantly weakened, and a checkerboard pattern may appear.  A temperature variation of about 10\% can induce spirals with order unity density perturbations, comparable to those generated by a thermal mass planet. We then study ring formation and find it is related to the coupling between azimuthal temperature variations and spirals outside the resonances. Such coupling leads to a radially varying angular momentum flux, which produces anomalous wave-driven accretion and forms dense rings separated by the wavelength of the waves. Finally, we speculate that spirals induced by temperature variations may contribute to disk accretion through non-linear wave steepening and dissipation. Overall, considering that irradiation determines the temperature structure of protoplanetary disks, the change of irradiation both spatially or/and temporarily may produce observable effects in protoplanetary disks, especially spirals and rings in outer disks beyond tens of AU.

\end{abstract}


\section{Introduction}

Substructures such as gaps, rings, spirals, and crescents are ubiquitous in protoplanetary disks (e.g. \citealt{Andrews2020, Bae2023, Benisty2023}). The kinematic properties of these structures have also been revealed by ALMA molecular line observations \citep{Pinte2023}. These structures may facilitate planet formation, e.g., by concentrating dust particles. Meanwhile, they may be induced by young planets and thus serve as signposts of young planets in disks. Although numerous physical mechanisms have been proposed to explain these substructures (see review by \citealt{Bae2023}), most involve gravitational (including planetary and stellar perturbers and disk self-gravity) and magnetic processes (such as zonal flows, magnetocentrifugal winds, and deadzone boundaries).  

In contrast, thermal perturbations have received comparatively little attention. Most studies to date on thermodynamic processes in disks focus primarily on instabilities within the disk \citep{Lesur2023}. These studies emphasize that the operation of these instabilities is highly sensitive to the background temperature structure \citep{Nelson2013, zhang24} and the cooling efficiency \citep{lin15}. However, the background disk is typically assumed to be axisymmetric with a uniform temperature in the azimuthal direction. 

Recent observations challenge this assumption of azimuthal symmetry in temperature.  Near-infrared scattered light observations (e.g. \citealt{marino15}) reveal that many protoplanetary disks, especially transitional disks, exhibit asymmetric shadows (see review by \citealt{Benisty2023}), which could be cast by misaligned inner disks or magnetospheric accretion columns around the star. The temperature could decrease inside these shadow, as evidenced by reductions in dust continuum emission in ALMA observations \citep{casassus15, ubeira19, safonov22, arce-tord23}, creating pressure imbalances and thus an azimuthal driving forces. Since any asymmetric perturbation can launch spiral density waves  (e.g. \citealt{Zhu2022}), shadows can therefore launch spiral waves. 

Two-dimensional numerical simulations with simplified heating/cooling treatments \citep{montesinos16, montesinos18, cuello19, su24} have revealed spirals in shadowed disks.  Recent three-dimensional radiation hydrodynamical simulations \citep{Zhang2024b}  confirm that spirals are robust features launched by shadows, and the spirals in the synthetic images resemble those observed in transitional disks. Besides spirals, rings and asymmetric disk features could also be induced by disk shadows \citep{su24, qian24,Ziampras2024}.
In this work, we develop a linear theory to study structure induced by asymmetric temperature perturbations in disks. Such temperature perturbations could arise not only from asymmetric disk shadows but also from local heating or cooling events, such as accreting planets \citep{Zhu2015,Cleeves2015} or shadows cast by circumplanetary material \citep{Montesinos2021,Muley2024}. We focus on orbiting temperature perturbations in this work and will study stationary temperature perturbations (in the inertial frame) in a subsequent paper (Zhu \etal in prep.).

We describe the numerical simulations in \S 2, and introduce the linear perturbation theory in \S 3. The results are presented in \S 4. After discussions in \S 5, we conclude the paper in \S 6.

\section{Numerical Simulations}

\begin{figure}[t!]
\includegraphics[trim=0mm 0mm 0mm 0mm, clip, width=3.3in]{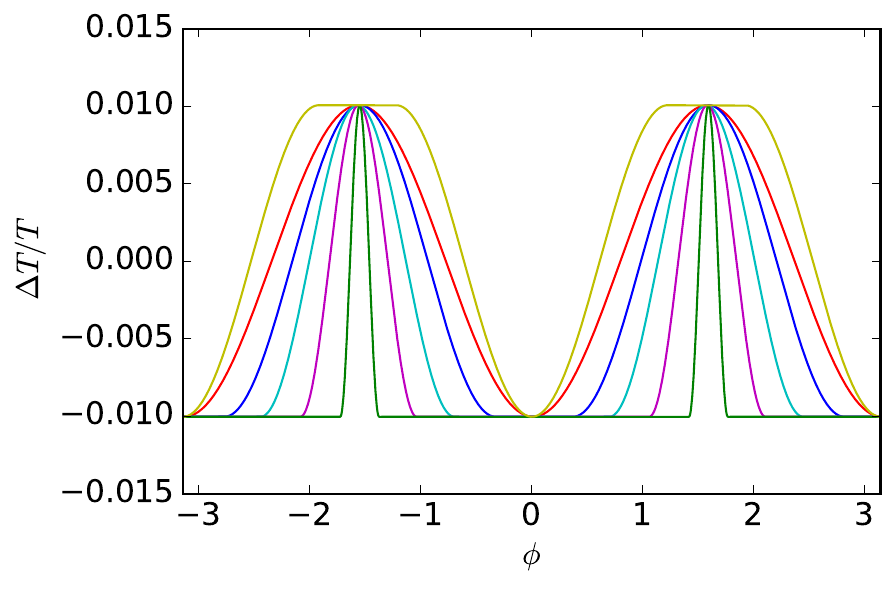}
\figcaption{Temperature perturbations with different azimuthal widths in simulations.
\label{fig:sizetemperature}}
\end{figure}

\begin{figure*}[t]
\includegraphics[trim=2mm 0mm 2mm -6mm, clip, width=7in]{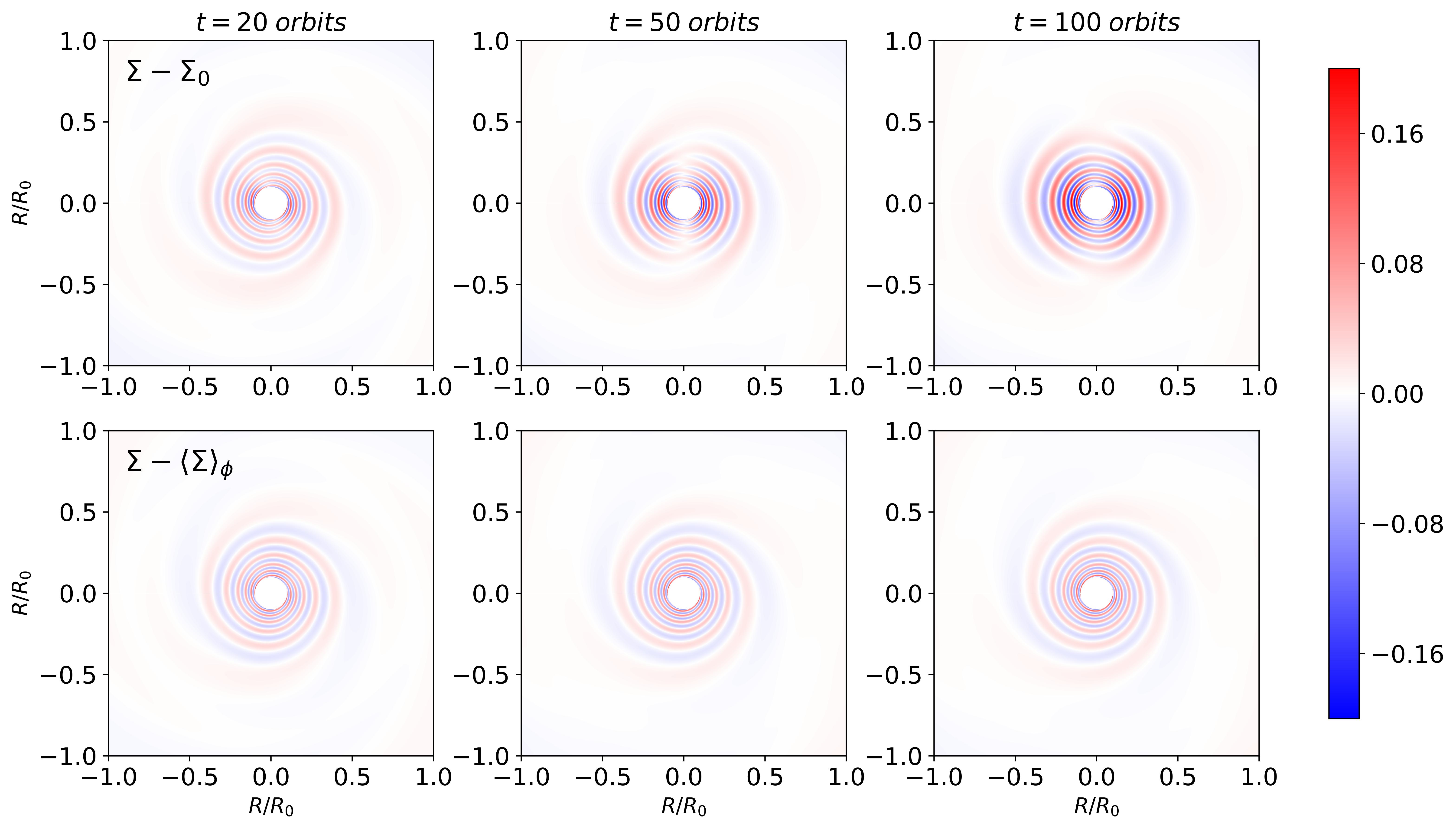}
\figcaption{Density perturbations from the direct numerical simulation with $f_{sh}=0.01$ at different times. The upper panels plot the density perturbation compared with the initial condition, which shows that spirals become rings. The lower panels only plot the non-axisymmetric density perturbation by subtracting the density with the azimuthally averaged surface density at that time. 
\label{fig:evoution}}
\end{figure*}

We have carried out two-dimensional ($R-\phi$) hydrodynamical simulations for disks with azimuthal temperature variations using the  Athena++ code \citep{Stone2020}. These direct numerical simulations allow us to investigate both the linear effects for shallow shadows and the nonlinear effects when the shadows are deep.

Our initial disk structure follows 
\begin{align}
\Sigma(R,\phi)&=\Sigma(R_0) (R/R_0)^{p}\\
T(R,\phi)&=T(R_0)(R/R_0)^{q}\label{eq:Tr}\,,
\end{align}
where $R_0$ is the corotation radius $R_p$ of the pattern. For the planetary case, $R_p$ corresponds to the planet's orbital distance; for the shadowed case, it is the corotation radius of the rotating shadow in the disk. We set $q=-1/2$, and  the disk aspect ratio $h\equiv c_s/v_{K}$ as 0.1 at $R_0$. To explore the effects of the surface density gradient, we consider values of $p$ equal to 0 (fiducial cases) and -1.

To allow the initial disk to reach a steady state, the shadows are not introduced until $t=10\times 2\pi/\Omega_0$. After this settling period, we introduce an azimuthal temperature perturbation that orbits the central star at a pattern speed $\Omega_p$ equal to the Keplerian speed $\Omega_K(R=R_0)$ at $R_0$.

Motivated by the shadow temperature profiles used in radiation-hydrodynamical simulations \citep{zhang24}, where the inner disk is perpendicular to the outer disk, we set up two symmetric shadow lanes separated by 180$^\circ$. We consider two azimuthal temperature levels for the irradiated temperature: the fully shadowed temperature $T_{low}$ and the fully irradiated temperature $T_{high}$. The radial profile of $T_{high}$ follows Equation \ref{eq:Tr}.  The corresponding sound speeds are $c_{s,T_{low}}$ and $c_{s,T_{high}}$, respectively.  The shadows are centered at 0 and $\pi$, while the fully irradiated regions are centered at $\pi$/2 and 3$\pi$/2 initially. The shadow half-width is $\phi_{T_{low}}$, and the fully irradiated region's half-width is $\phi_{T_{high}}$. Within the half-widths around both the shadowed and the fully irradiated regions, the temperature is constant. At each radius, transition regions exist between the half-widths of the shadowed and irradiated regions. The transitions from shadowed to fully irradiated regions are smoothly connected using sinusoidal functions:
\begin{equation}
c_{s,irr} = c_{s,T_{low}} + (c_{s,T_{high}}- c_{s,T_{low}})\textrm{sin}^2\Bigg(\frac{\pi}{2}\frac{\phi-\phi_{start}}{\phi_{end}-\phi_{start}}\Bigg)\,,
\end{equation}
where $\phi_{start}$ and $\phi_{end}$ are the starting and end points of the transition regions. Similarly, the transitions from fully irradiated to shadowed regions are connected using cosine functions:
\begin{equation}
c_{s,irr} = c_{s,T_{low}} + (c_{s,T_{high}}- c_{s,T_{low}})\textrm{cos}^2\Bigg(\frac{\pi}{2}\frac{\phi-\phi_{start}}{\phi_{end}-\phi_{start}}\Bigg)\,.
\end{equation}
In our fiducial case, we set $\phi_{T_{low}}=\phi_{T_{high}}=0^o$, so that the temperature perturbation corresponds to an $m=2$ Fourier mode. Except for the simulation in \S 4.5, all our simulations use $c_{s,T_{low}}/c_{s,T_{high}}=0.99$ at all radii, which corresponds to $f_{sh}=0.01$ in the linear analysis later. As will be presented in \S 4.4, we have also set up a variety of azimuthal temperature profiles with $\phi_{T_{low}}=20^o$, $40^o$, $60^o$, $80^o$, and $\phi_{T_{high}}=0^o$, $20^o$. The azimuthal temperature profiles for all $c_{s,T_{low}}/c_{s,T_{high}}=0.99$ cases are shown in Figure \ref{fig:sizetemperature}.

We apply a simple orbital cooling scheme to relax the disk temperature to $T_{irr}$:
\begin{equation}
\frac{dE}{dt}=-\frac{E-E_{irr}}{t_{cool}}\,,
\end{equation} 
where $\Sigma$ and $E$ are the disk surface density and the internal energy per unit area, and $E_{irr}\equiv \Sigma c_{s,irr}^2/(\gamma-1)$ is the internal energy per unit area with the irradiation temperature. The cooling time $t_{cool}$ can be expressed in the dimensionless form as $\beta=t_{cool}\Omega(R)$. We vary $\beta$ as $10^{-6}$, $10^{-2}$, $10^{-1}$, $1$, $10$, $100$. The case with $\beta=10^{-6}$ is equivalent to the locally isothermal case since the cooling time is shorter than the simulation timestep.

Our simulation domain extends from $R=0.1$ to 10$R_0$ with 750 grid points in the radial direction and 1024 grid points in the azimuthal direction. The grid spacing is uniform in  logarithmic space radially and uniform in  linear space azimuthally. We run the simulations for 100 orbits at $R_0$. Unless stated otherwise, the presented simulation results correspond to the end of the simulation at 100 orbits at $R_0$. 
 Periodic boundary conditions are applied in the azimuthal direction. 
 To limit wave reflection, density and velocity at the radial boundaries are set to the initial conditions during each timestep. The temperatures at the boundaries are also reset at each timestep. The reset temperatures have the azimuthal variations that are identical to those in the disk. In other words, the boundaries are also subject to shadows. We have tried to reset the temperature to the initial condition without the azimuthal variation. This makes little difference except for the slow cooling cases that will be discussed in \S 4.3.

Figure \ref{fig:evoution} shows the density evolution for our fiducial locally isothermal case with $f_{sh}=0.01$. The upper row shows the density perturbation above a constant value. Spirals are launched initially. At 50 orbits, the spiral arms start to break at $\theta=\pi/2$, $3/2\pi$ where the temperature is the highest.  Then, they seem to reconnect and finally form rings (100 orbits). This secular evolution is due to the disk accretion rate that varies radially. 

The bottom row in Figure \ref{fig:evoution} shows the azimuthal density perturbation that is derived by subtracting the density with the azimuthally averaged density at each radius. The resulting spiral patterns are quite stable,  indicating that the spirals (linear features as shown below) persist even when rings are forming. Considering that disk accretion is a non-linear secular process and is more difficult to study, we start with linear analysis which may shed light on the non-linear evolution.

\section{Linear Perturbation Theory}
To understand the spirals and rings,  we have carried out the linear analysis for an inviscid disk governed by the Euler equations:
\begin{eqnarray}
\frac{\partial \rho}{\partial t} + \nabla\cdot\rho {\bf v} &=& 0 \label{eq:mass}\\
\frac{\partial {\bf v}}{\partial t} +({\bf v}\cdot\nabla) {\bf v} +\frac{\nabla p}{\rho} &=& -\nabla\Phi_{ext} \label{eq:vel}
\end{eqnarray}
where $\Phi_{ext}$ represents any external potential acting on the disk (e.g., gravitational potential from a star or/and a planet). 
We adopt a locally isothermal equation of state as $p=\rho c_{s}^2$ with $c_{s}$ being the local isothermal sound speed. We note that although it is possible to develop a linear theory considering a non-zero $\beta$ cooling \citep{Miranda2020}, the equations become significantly more complicated. Furthermore, unlike the planet-disk interaction problem in  \cite{Miranda2020}, the driving force due to the shadow is significantly reduced when $\beta>0.1$, as shown in \S 4.3.  Thus, the slow cooling case is less relevant, and we focus on the locally isothermal cases here. 
We decompose the temperature into axisymmetric and non-axisymmetric components:
\begin{equation}
 T(R,\phi,z,t)=T_{0}(R,z)+T_{v}(R,\phi,z,t)\,, 
\end{equation}
where $T_0$ is the axisymmetric temperature profile, and $T_{v}$  represents temperature variations due to shadows or other asymmetric heating/cooling processes. Correspondingly, the pressure $p=\rho c_{s}^2=\rho c_{s,T0}^2+\rho c_{s,Tv}^2\equiv p_{T0}+p_{Tv}$. $p_{T0}$ is the pressure that is calculated using $\rho$ and the axisymmetric temperature $T_0$. $p_{Tv}$ is the pressure calculated with $\rho$ and the perturbed temperature $T_v$, and thus $c_{s,Tv}^2=T_{v}R_{sp}$ with the specific gas constant $R_{sp}$. To quantify the amplitude of the temperature perturbation, we define $f_{sh}$ as the ratio between the maximum temperature variation and the background temperature at each radius and $z$:
\begin{equation}
f_{sh}(R,z)=\frac{max_{\phi} (T_{v}(R,\phi,z,t))}{T_{0}(R,z)}\,.
\end{equation}
In this work, we assume that $T_{v}(R,\phi,z,t)$ varies with time, orbiting around the star at the orbital frequency $\Omega_p$.
 
Substituting the decomposed pressure into the momentum equation (Equation \ref{eq:vel}), we obtain:
\small
\begin{eqnarray}
\frac{\partial {\bf v}}{\partial t} +({\bf v}\cdot\nabla) {\bf v} +\frac{\nabla p_{T0}}{\rho} &=& -\frac{\nabla p_{Tv}}{\rho}  -\nabla\Phi_{ext}\,\label{eq:velpts}\\
&=&-\frac{c_{s,Tv}^2\nabla \rho}{\rho} -\nabla({c_{s,Tv}^2} + \Phi_{ext})\,,\label{eq:vel2}
\end{eqnarray}
\normalsize
where the second equation has used $\nabla p_{Tv}=c_{s,Tv}^2\nabla \rho + \rho \nabla c_{s,Tv}^2$. 

If the temperature variations are small, the zeroth order equations have the traditional axisymmetric solution of $\rho_0$ and ${\bf v_0}$. We could separate quantities as
$\rho=\rho_0+\rho_1$, ${\bf v}={\bf v_0}+{\bf v_1}$, and $p_{T0}=p_{T0,0}+p_{T0,1}$, where $p_{T0,0}$ is the unperturbed pressure calculated using $T_0$ and $\rho_0$, and $p_{T0,1}$ is the perturbed pressure calculated using $T_0$ and $\rho_1$. Then, the linearized continuity and momentum equations for
perturbed quantities are
\begin{align}
&\frac{\partial \rho_1}{\partial t} +  \nabla\cdot(\rho_0 {\bf v_1}+\rho_1 {\bf v_0})  = 0 \label{eq:mass}\\
&\frac{\partial {\bf v_1}}{\partial t} +({\bf v_0}\cdot\nabla) {\bf v_1}+({\bf v_1}\cdot\nabla) {\bf v_0}  \nonumber\\
&= \frac{\rho_{1}\nabla p_{T0,0}}{\rho_0^2}-\frac{\nabla p_{T0,1}}{\rho_0}-\frac{c_{s,Tv}^2\nabla \rho_0}{\rho_0} -\nabla({c_{s,Tv}^2} + \Phi_{ext,1}) \,,\label{eq:velper}
\end{align}
where $\Phi_{ext,1}$ is the potential from an additional perturber besides the star (e.g. a planet). 
We can see that the temperature variations in the disk act like driving accelerations, shown as the first term on the right-hand side of Equation \ref{eq:velpts} or the third and fourth terms on the right side of Equation \ref{eq:velper}. 
If the background density gradient is zero, $c_{s,Tv}^2\nabla\rho_0/\rho_0$ term in Equation \ref{eq:velper} is zero, and the temperature perturbations effectively introduce an additional potential in Equation \ref{eq:velper}:
\begin{equation}
\Phi_{T,1}=c_{s,Tv}^2\,.
\end{equation}
In this uniform density case, this is identical to the perturbation by an external potential as in the planet-disk interaction problem.  Although the temperature variation  can be mathematically treated similarly to an external potential ($\nabla c_{s,Tv}^2$ in Equation \ref{eq:vel2}), the disk's total angular momentum is conserved (if there is no advection through the boundaries and $\phi_{ext}$ is only from the central star), since the pressure is an internal force.
This is different from the planet-disk interaction problem where the disk's total angular momentum can be changed due to the planetary torque.

We decompose the temperature and potential perturbations into azimuthal Fourier components:
\begin{align}
c_{s,Tv}^2=\sum_{m=0}^{\infty}{\rm Re}\left[\Upsilon_{m}(R,z)e^{im(\phi-\Omega_p t)}\right]\,,\\
\Phi_{ext,1}=\sum_{m=0}^{\infty}{\rm Re}\left[\Phi_{m}(R,z)e^{im(\phi-\Omega_p t)}\right]\,,
\end{align}
where $m$ is the azimuthal mode number, and $\Omega_p$ is the pattern speed of the perturbation. 
For a two-dimensional disk in cylindrical coordinates  ($R$, $\phi$), we integrate $\rho$ and $p$ vertically to obtain surface density $\Sigma$ and vertically integrated pressure $P$. We assume that the velocity and temperature structure is height independent, so that $\rho$ and $p$ in Equations \ref{eq:mass} and \ref{eq:velper} can be replaced with $\Sigma$ and $\Sigma c_{s}^2$. We consider the perturbations of the form:
\begin{equation}
\delta x(R,\phi,t)= \sum_{m=0}^{\infty}{\rm Re}\left[\delta x_m(R)e^{im(\phi-\Omega_p t)}\right]\,.
\end{equation}
For each Fourier component (we will drop the subscript $m$ for discussing one component), the linearized equations become: 
\begin{align}
-i\tilde{\omega}\Sigma_1&+\frac{1}{R}\frac{d}{d R}\left(R\Sigma_0 v_{R,1}\right)+\frac{i m \Sigma_0}{R} v_{\phi,1}=0 \label{eq:sigmaf}\,,\\
-i\tilde{\omega}v_{R,1}&-2\Omega v_{\phi,1}=-\frac{1}{\Sigma_0}\frac{d }{d R} P_{T0,1}+\frac{1}{\Sigma_0^2}\frac{d P_{T0,0}}{dR}\Sigma_{1}\nonumber\\
&-\frac{\Upsilon_m}{\Sigma_0}\frac{d \Sigma_0}{dR}-\frac{d}{d R}\left(\Phi_m+\Upsilon_m\right)\label{eq:vrf}\\
-i\tilde{\omega}v_{\phi,1}&+\frac{\kappa^2}{2\Omega}v_{R,1}=-\frac{im}{R}\left(\frac{P_{T0,1}}{\Sigma_0}+\Upsilon_m+\Phi_m\right) \label{eq:vphif}
\end{align}
where $\tilde{\omega}\equiv m(\Omega_p-\Omega)$, and 
\begin{equation}
\kappa^2\equiv\frac{2\Omega}{R}\frac{d}{dR}\left(R^2\Omega\right)\,.
\end{equation}
Thus, besides all the normal terms in the perturbation equations, the temperature variation leads to an additional equivalent potential ($\Upsilon_m$) and a  force term ($d\Sigma_0/dR$ term) in Equation \ref{eq:vrf}.
Again, if the background surface density is constant with radius, $d\Sigma_0/dR$ vanishes, and the temperature perturbation acts solely through the equivalent potential $\Upsilon_m$, analogous to $\Phi_m$. 

The velocity perturbations can be derived from Equations \ref{eq:vrf} and \ref{eq:vphif}:
\begin{align}
v_{R,1}=&\frac{i}{D}\left[\left(\tilde{\omega}\frac{d}{d R}-\frac{2m\Omega}{R}\right)\left(\frac{P_{T0,1}}{\Sigma_0}+\Phi_m+\Upsilon_m\right)\right.\nonumber\\
&\left.+\tilde{\omega}\left(\frac{\Upsilon_m}{L_\Sigma}-\frac{1}{L_T}\frac{P_{T0,1}}{\Sigma_0}\right)\right]\,,\label{eq:vr1}\\
v_{\phi,1}=&\frac{1}{D}\left[\left(\frac{\kappa^2}{2\Omega}\frac{d}{d R}-\frac{m\tilde{\omega}}{R}\right)\left(\frac{P_{T0,1}}{\Sigma_0}+\Phi_m+\Upsilon_m\right)\right.\nonumber\\
&\left.+\frac{\kappa^2}{2\Omega}\left(\frac{\Upsilon_m}{L_\Sigma}-\frac{1}{L_T}\frac{P_{T0,1}}{\Sigma_0}\right)\right]\label{eq:vp1}\,,
\end{align}
where 
\begin{align}
D=\kappa^2-\tilde{\omega}^2\,,\\
\frac{1}{L_T}=\frac{d\; {\rm ln} c_{s,T0}^2 }{dR}\,,\\
\frac{1}{L_\Sigma}=\frac{d \;{\rm ln} \Sigma_{0}^2 }{dR}\,.
\end{align}
Substituting $v_{R,1}$ and $v_{\phi,1}$ back into Equation \ref{eq:sigmaf} yields a second-order ODE for $P_{T0,1}/\Sigma_0$:
\begin{equation}
\frac{d^2}{dR^2}\left(\frac{P_{T0,1}}{\Sigma_0}\right)+C_1\frac{d}{dR}\left(\frac{P_{T0,1}}{\Sigma_0}\right) + C_0\frac{P_{T0,1}}{\Sigma_0}=\Psi_m\,,\label{eq:ODE}
\end{equation}
with
\begin{align}
C_1=&\frac{d}{dR}\left({\rm ln}\;\frac{R\Sigma_0}{D}\right)-\frac{1}{L_T}\,,\\
C_0=&-\frac{2m\Omega}{R\tilde{\omega}}\left[\frac{1}{L_T}+\frac{d}{dR}\left({\rm ln}\; \frac{\Sigma_0\Omega}{D}\right)\right]\nonumber\\
&-\frac{1}{L_T}\frac{d}{d R} \left({\rm ln}\;\frac{R\Sigma_0}{DL_T}\right)-\frac{m^2}{R^2}-\frac{D}{c_{s,T0}^2}\,,\\
\Psi_m=&
-\frac{d^2}{dR^2}\left(\Phi_m+\Upsilon_m\right)-\frac{d}{dR}\left({\rm ln\; \frac{R\Sigma_0}{D}}\right)\frac{d}{dR}\left(\Phi_m+\Upsilon_m\right)\nonumber\\
&+\left(\frac{2m\Omega}{\tilde{\omega}R}\frac{d}{dR}\left({\rm ln}\frac{\Omega\Sigma_0}{D}\right)+\frac{m^2}{R^2}\right)\left(\Phi_m+\Upsilon_m\right)\nonumber\\
&-\frac{d}{dR}\left(\frac{\Upsilon_m}{L_{\Sigma}}\right)-\left(\frac{d}{dR}\left({\rm ln}\;\frac{R\Sigma_0\tilde{\omega}}{D}\right)+\frac{m\kappa^2}{R\tilde{\omega}2\Omega}\right)\left(\frac{\Upsilon_m}{L_{\Sigma}}\right)\,.\label{eq:Phim}
\end{align}
Comparing these equations with those for the planet-disk interaction problem (e.g., Equations (17)-(19) in \cite{Miranda2020}), we can see that the only additional terms are those involving $\Upsilon_m/L_{\Sigma}$ in the source term $\Psi_m$ (last line of Equation \ref{eq:Phim}). This highlights the analogy between temperature-induced perturbations and those caused by external potentials.

Singularity occurs at the Lindblad resonances where $D=0$. Both analytical and numerical techniques can be used to treat this singularity \citep{GoldreichTremaine1979}. 
We follow the numerical procedure in \cite{Korycansky1993,Miranda2019} to solve this equation with the outgoing WKB wave boundary conditions (the details are given in the Appendix of \citealt{Miranda2019}). These conditions  allow waves to leave the domain freely and no incoming waves enter the domain.
To simplify the problem, we further assume that  the amplitude of the temperature variation is a constant fraction of the background temperature  with
$\Upsilon_m(R)= f_{sh,m }c_{s,T0}(R)^2$, the same as our direct numerical simulations. 

By solving the second-order ODE with appropriate boundary conditions, we can determine the disk's response to temperature perturbations. We will confirm the results from linear calculations with those from direct numerical simulations.

\section{Results}
\subsection{Comparison between Linear Analysis and Numerical Simulations}

\begin{figure}[t]
\includegraphics[trim=2mm 0mm 2mm -6mm, clip, width=3.4in]{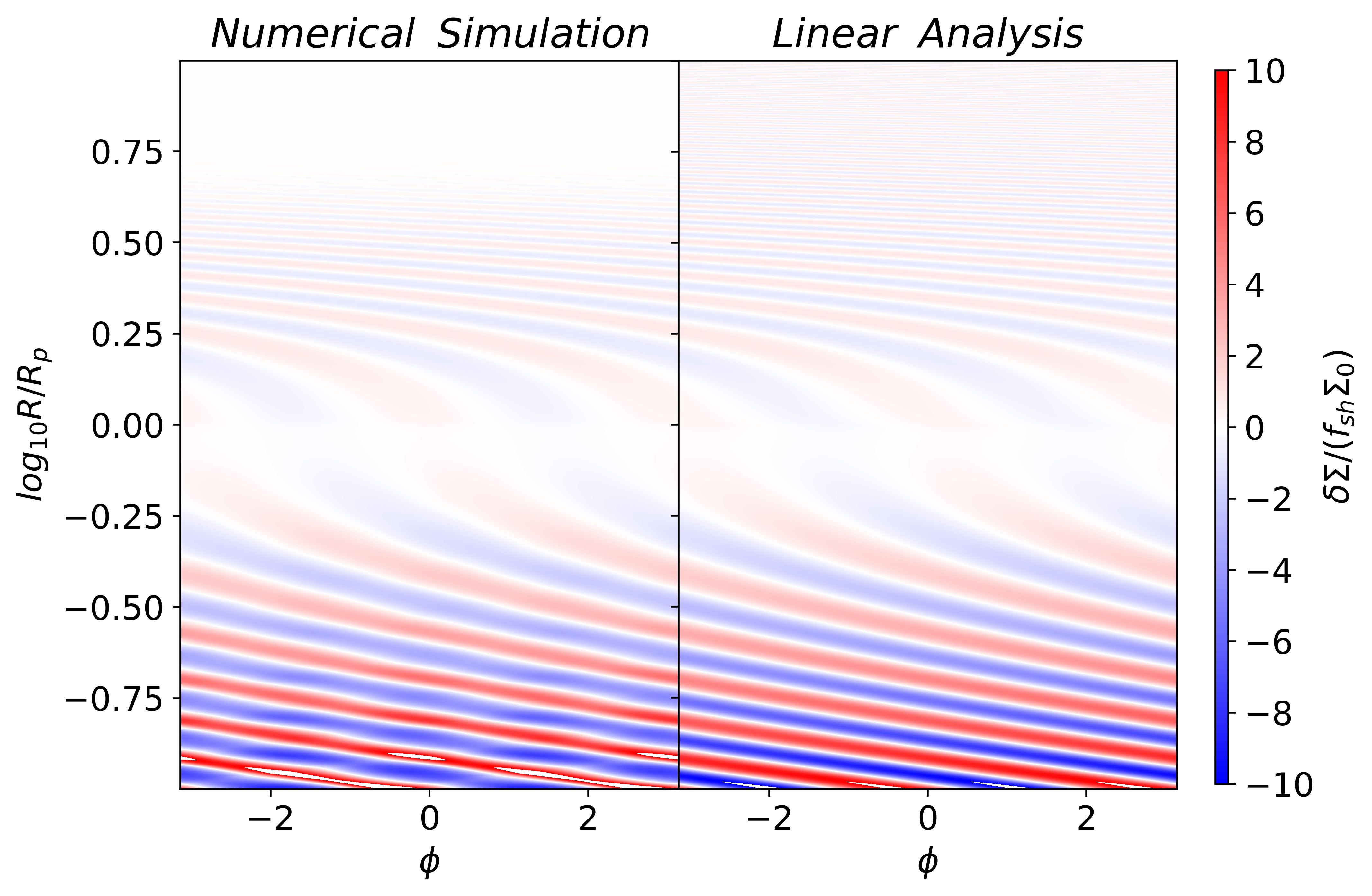}
\figcaption{Density perturbations from the direct numerical simulation at the end of the simulation (left panel) and the linear analysis (right panel). An m=2 shadow temperature is imposed. The density perturbation is normalized to $f_{sh}\Sigma_0$.
\label{fig:Twodlinear}}
\end{figure}

\begin{figure}[h]
\includegraphics[trim=0mm 0mm 0mm 0mm, clip, width=3.3in]{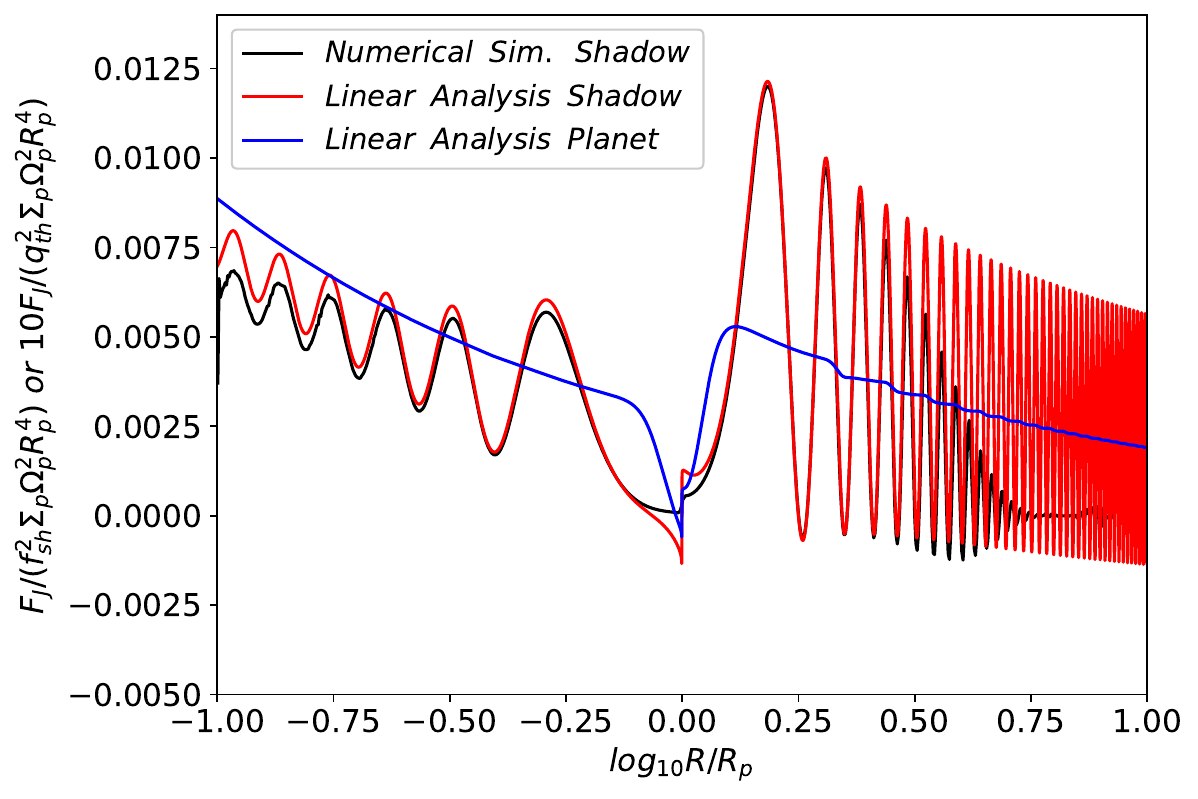}
\figcaption{Black and red curves: the angular momentum flux for a disk with an m=2 shadow. The blue curve: the angular momentum flux for a smooth disk having a planet. To show values in the same scale, we have normalized the angular momentum flux in the planetary case by 1/10th of $q_{th}^2\Sigma_p\Omega_p^2R_p^4$.
\label{fig:torque}}
\end{figure}

Using our fiducial setup, we have performed both linear analysis and direct numerical simulations to investigate the disk's response to asymmetric temperature perturbations. In the linear analysis, we generate maps of density perturbations using the derived linear modes, as shown in Figure \ref{fig:Twodlinear}. We  also calculate the angular momentum fluxes from both approaches
\begin{equation}
F_J(R)=R^2\int \Sigma v_{R}\delta v_{\phi}d\phi\,,
\end{equation}
where $\delta v_{\phi}=v_{\phi}-<v_{\phi}>_{\phi}$, as shown  in Figure \ref{fig:torque}. Overall, both Figures demonstrate excellent agreement between the two approaches. The only notable discrepancy occurs in the outer regions of the disk, where the spirals become very tightly wound, leading to significant numerical dissipation in the simulations. 

In the linear regime, the amplitudes of the primitive quantities ($\delta\Sigma$, $\delta v$) are proportional to the amplitude of the driving temperature perturbations, characterized by $f_{sh}$. Accordingly, in Figure \ref{fig:Twodlinear}, we scale $\delta \Sigma$ with $f_{sh}\Sigma_{0}$. Consequently, the angular momentum flux scales with $f_{sh}^2$. Therefore, as shown in Figure \ref{fig:torque},  $F_J$ should be normalized by $f_{sh}^2\Sigma_p\Omega_p^2R_p^4$. The subscript $p$ indicates the quantities at the corotating radius $R_p$, where the Keplerian angular velocity equals the pattern speed $\Omega_p$.

For comparison, we have also plotted in Figure \ref{fig:torque} the angular momentum flux excited by a planet  embedded in the disk. It is derived from the linear analysis with the locally isothermal equation of state, following \cite{Miranda2020}. The planet potential is smoothed over a scale of 0.6 disk scale heights. We define $q_{th}\equiv M_p/M_{th}$ where $M_{th}$ is $h^3 M_*$, the so-called ``thermal mass''. The angular momentum flux induced by the planetary perturber is smooth and increases toward smaller radii. In contrast, for the $m$=2 temperature perturbation case, the angular momentum flux varies with radius, although it also increases toward smaller radii.

The increase of the angular momentum flux (AMF) towards smaller radii can be understood within the framework of linear theory. If only small amplitude waves are transporting the AMF, the AMF becomes
\begin{equation}
F_J(R)=R^2\Sigma(R)\int \delta v_{R}(R,\phi)\delta v_{\phi}(R,\phi)d\phi\,,
\end{equation}
where $\delta v_{R}$ and $\delta v_{\phi}$ are the radial and azimuthal components of the velocity perturbations, respectively. 
For a particular azimuthal mode $m$ in the linear analysis, the AMF can be simplified to
\begin{equation}
F_J^{m}(R)=\pi R^{2}\Sigma_0 {\rm Re}[v_{R,1}v_{\phi,1}^{*}]=\frac{\pi}{2} R^{2}\Sigma_0 (v_{R,1}v_{\phi,1}^{*}+v_{R,1}^*v_{\phi,1})\,,\label{eq:fjm}
\end{equation}
where the asterisk denotes complex conjugation.
 
By substituting Equations \ref{eq:vr1} and \ref{eq:vp1} into Equation \ref{eq:fjm}, we obtain
\small
\begin{align}
F_J^m=&\frac{\pi R m \Sigma}{D}{\rm Im}\left[\left(\frac{P_{T0,1}}{\Sigma_0}+\Phi_m+\Upsilon_m\right)\frac{d}{dR}\left(\frac{P_{T0,1}}{\Sigma_0}^*+\Phi_m^*+\Upsilon_m^*\right)\right.\,\nonumber\\
&\left.+\left(\frac{P_{T0,1}}{\Sigma_0}^*+\Phi_m^*\right)\frac{\Upsilon_m}{L_\Sigma}-\left(\Phi_m^*+\Upsilon_m^*\right)\frac{1}{L_T}\frac{P_{T0,1}}{\Sigma_0}\right]\,,\label{eq:Fjm}
\end{align}
\normalsize
where Im denotes the imaginary part of the complex number.
We can decompose $F_J^m$ into $F_J^m=F_{J,h}^m+F_{J,r}^m$, where
\begin{equation}
F_{J,h}^m=\frac{\pi R m \Sigma}{D}{\rm Im}\left[\left(\frac{P_{T0,1}}{\Sigma_0}\right)\frac{d}{dR}\left(\frac{P_{T0,1}}{\Sigma_0}^*\right)\right]\,,\label{eq:fjh}
\end{equation}
and $F_{J,r}^m$ includes the remaining terms that are related to the potential and temperature perturbations. 

Following \cite{Miranda2020}, we can derive that, for a freely propagating wave (where $\Phi_m$ and $\Upsilon_m$ are zero in  Equation \ref{eq:ODE}), 
\begin{equation}
\frac{d}{dR}\left(\frac{F_{J,h}^m}{c_{s,T0}^2}\right)=0\,.\label{eq:fjht}
\end{equation}
This implies that $F_{J,h}^m/c_{s,T0}^2$ is conserved, and since the background temperature increases toward smaller radii (with $q=-1/2$ in our simulations), 
the angular momentum flux increases toward smaller radii. In the case of a planetary perturber,
the planetary potential decreases rapidly with distance from the planet, so that $\Phi_m$ in Equation \ref{eq:Fjm} becomes negligible far from the planet, and $F_{J,r}^m\sim 0$. Thus, the angular momentum flux follows Equation \ref{eq:fjht}, resulting in a smooth increase toward smaller radii. 

On the other hand, in the case with disk shadows, $\Upsilon_m$ is non-zero at all radii. Consequently, $\Upsilon_m$ can couple with the $P_{T0,1}/\Sigma_0$ terms in Equation \ref{eq:Fjm}, leading to a radially varying angular momentum flux. In the WKB limit, $P_{T0,1}/\Sigma_0$ has a wavenumber $k$ that follows the dispersion relation
\begin{equation}
m^2(\Omega_p-\Omega)^2=\kappa^2+c_{s}^2 k^2\,.
\end{equation}
This coupling causes variations in $F_J$ with the wavenumber $\sim k$, as observed in Figure \ref{fig:torque}. This effect is somewhat analogous to the additional torque contribution in the planetary companion case, where distant spirals can couple with the planetary potential at its azimuthal angle \citep{Cimerman2024}. 

\subsection{Perturbation Amplitude}
\begin{figure*}[t!]
\includegraphics[trim=5mm 7mm 0mm 0mm, clip, width=7.in]{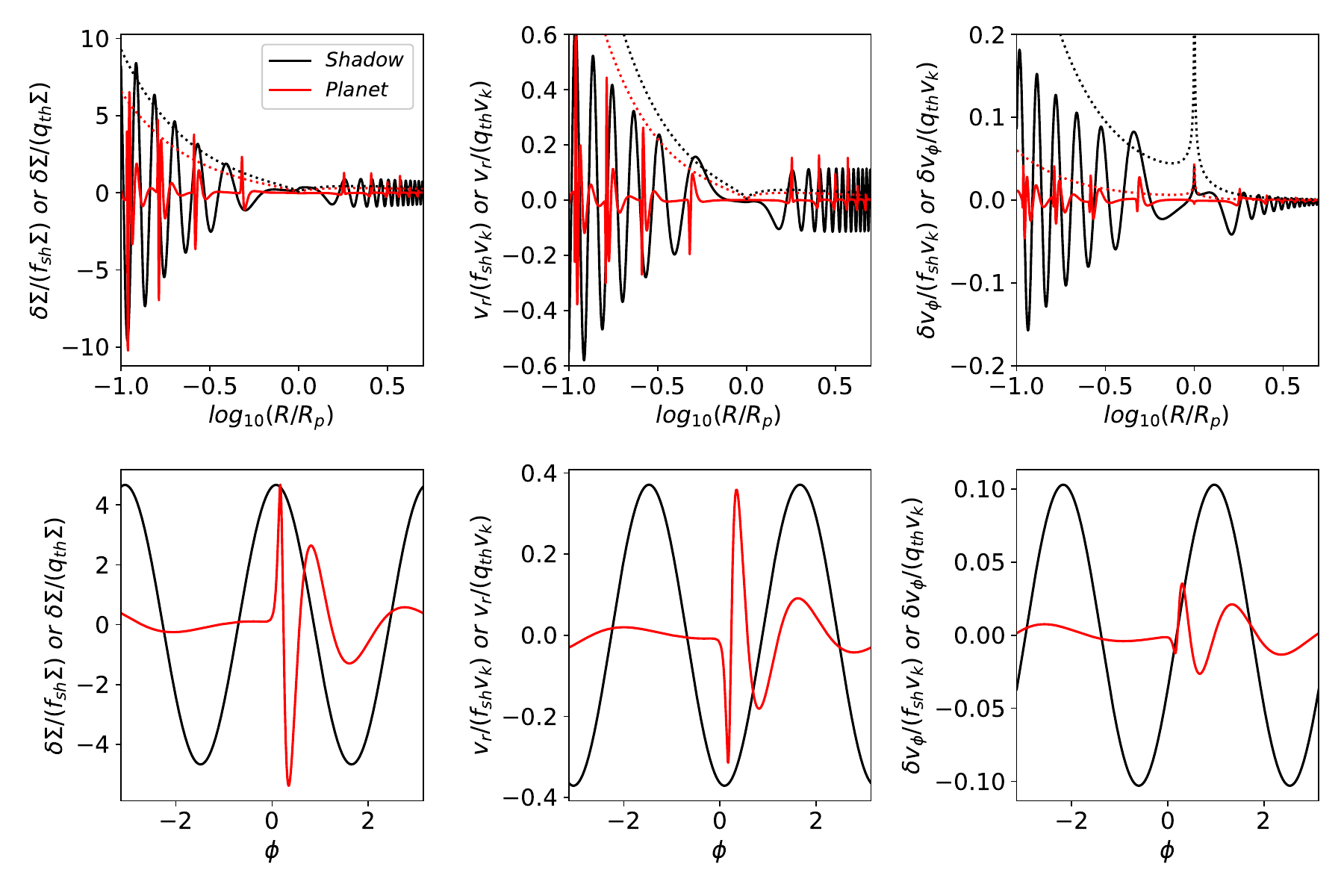}
\figcaption{The density, radial velocity, and azimuthal velocity perturbations (from left to right panels) along the radial (upper panels) and azimuthal directions (lower panels) for the shadowed disk case (black curves) and the planet-disk interaction case (red curves). Quantities are from cuts at $\phi=\pi$ in the upper panels and $R=0.2 R_p$ in the lower panels. Both cases are from linear calculations. The dotted curves in the $\delta\Sigma$ panels are from the simple scaling from Equations \ref{eq:deltasigmap} (red curves) and \ref{eq:deltasigmat} (black curves), while the dotted curves in the velocity panels are from Equations \ref{eq:vr} and \ref{eq:vphi}.
\label{fig:linearcuts}}
\end{figure*}

Protoplanetary disk observations towards spiral arms, including both near-IR scattered light observations and ALMA CO kinematics observations, can constrain the amplitudes of the density and velocity perturbations within the spiral arms. Here, based on the linear analysis, we will provide some simple relationships between the spirals' perturbations and the azimuthal temperature  variations, which may be helpful for interpreting observations in future. From the linear analysis, the density and velocity perturbations along both radial and azimuthal directions 
are shown in Figure \ref{fig:linearcuts}. To plot both the shadowed and planetary cases on the same scale, we normalize the results with $f_{sh}$ for the shadowed case and $q_{th}$ for the planetary case. 

Even without directly solving the master equation, we can estimate the amplitude of the waves using the angular momentum equation. 
In the WKB limit, Equation \ref{eq:fjh} can be approximated as
\begin{equation}
F_{J,h}^m\approx \frac{Rc_{s,T0}^3}{|\Omega-\Omega_p|}\frac{\delta\Sigma^2}{\Sigma_0}\,,
\end{equation}
so that, for a freely propagating wave,
\begin{equation}
F_{J,h}^m\frac{c_{s,p}^2}{c_{s,T0}^2}=\frac{Rc_{s,T0}c_{s,p}^2}{|\Omega-\Omega_p|}\frac{\delta\Sigma^2}{\Sigma_0}=const\,, \label{eq:deltaS}
\end{equation}
where $c_{s,p}$ is the sound speed at the corotating radius $R_p$.  As discussed earlier, the coupling between $\Upsilon_m$ and $P_{T0,1}$ leads to additional variations in $F_J$ in the shadowed case. 
Nevertheless, we can still use Equation \ref{eq:deltaS} to estimate the density perturbations induced by either planets or temperature variations, as long as the net angular momentum flux ($F_{J,h}^m$) is larger or comparable to the fluctuating angular momentum flux ($F_{J,r}^m$). 

For planet-disk interactions, \cite{Goldreich1980} showed the torque due to the $m$-th harmonic is proportional to $m^2\Sigma_p/\Omega_p^2\times(GM_p/R_p)^2$, where $\Sigma_p$ is the surface density at $R_p$. Due to the torque cut-off effect, modes with $m>h^{-1}$ have diminishing contributions to the total torque. Summing over all modes up to $m\sim h^{-1}$ leads to an overall scaling with $h^{-3}$. Specifically, \cite{Goldreich1980} derived that the one-sided angular momentum flux is
\begin{equation}
 F_{J, planet}\sim 0.93 \Sigma_p R_{p}^4 \Omega_p^2 q^2 h_p^{-3}\label{eq:fplanet}\,,
\end{equation}
where $q=M_p/M_*$ is the planet-to-star mass ratio, and $h_p$ is the disk's aspect ratio at the planet's location.
With a smoothed potential, \cite{Miranda2020} derived that
\begin{equation}
 F_{J, planet}\sim 0.3 \Sigma_p R_{p}^4 \Omega_p^2 q^2 h_p^{-3}\label{eq:fplanet}\,,
\end{equation}
valid around the corotation radius.
By equating Equations \ref{eq:deltaS} and \ref{eq:fplanet}, we derive an expression for the relative density perturbation:
\begin{equation}
\frac{\delta \Sigma}{\Sigma(R)}=g_P \frac{M_p}{M_{th}}\left[\frac{\Sigma_p}{\Sigma_0(R)}\frac{R_p}{R}\frac{c_{s,p}}{c_{s,T0}(r)}\frac{|\Omega(r)-\Omega_p|}{\Omega_p}\right]^{1/2}\,.\label{eq:deltasigmap}
\end{equation}
where $g_P$ is a fudge factor. Based on Figure \ref{fig:linearcuts}, we choose $g_p=0.5$.

For the shadowed case, the driving ``potential'' is $c_{s,Tv}^2$ instead of $GM_p/R$. The velocity perturbations should scale with $ f_{sh,p} c_{s,p}^2/v_{p}$, where $f_{sh,p}$ is the fractional temperature perturbation at the pattern's corotation radius. Thus, the angular momentum flux should  scale with $f_{sh,p}^2\Sigma_p c_{s,p}^4/\Omega_{p}^2$.
Based on Figure \ref{fig:torque}, we find
\begin{equation}
F_{J, shadow}\sim20 f_{sh,p}^2\Sigma_p c_{s,p}^4 \Omega_p^{-2}\label{eq:fshadow}\,.
\end{equation}
Equating Equations \ref{eq:deltaS} and \ref{eq:fshadow}, we obtain
\begin{equation}
\frac{\delta \Sigma}{\Sigma(r)}=g_T f_{sh}\left(20 h_p\frac{\Sigma_p}{\Sigma_0}\frac{R_p}{R}\frac{c_{s,p}}{c_{s,T0}(R)}\frac{ |\Omega(r)-\Omega_{p}|}{\Omega_p}\right)^{1/2}\,,\label{eq:deltasigmat}
\end{equation}
where $g_T$ is a fudge factor. Based on Figure \ref{fig:linearcuts}, we also choose $g_T=0.5$.
To estimate the velocity perturbations, we apply the WKB approximation to Equations \ref{eq:vr1} and \ref{eq:vp1}. When the spirals are far from the excitation region, $|kR|>>1$, the $d/dR$ terms dominate, yielding
\begin{align}
v_{R,1}&=-k\frac{\tilde{\omega}}{D}\frac{\Sigma_1 c_{s,T0}^2}{\Sigma(R)}\sim \frac{\Sigma_1 c_{s,T0}}{\Sigma(R)}\label{eq:vr}\\
v_{\phi,1}&=-k\frac{i \kappa^2}{2\Omega D}\frac{\Sigma_1 c_{s,T0}^2}{\Sigma(R)}\sim\frac{i \Omega}{2 \tilde{\omega}}\frac{\Sigma_1 c_{s,T0}}{\Sigma(R)}\label{eq:vphi}
\end{align}
for freely propagating waves, where we have used $\tilde{\omega}\sim kc_{s,T0}(r)$, $D\sim-\tilde{\omega}^2$, and $\kappa^2=\Omega^2$. 

Equations \ref{eq:deltasigmap}, \ref{eq:deltasigmat}, \ref{eq:vr}, and \ref{eq:vphi} are plotted as dotted curves in Figure \ref{fig:linearcuts}.  For the planetary case, we choose $m$=$1/h$=10, while for the shadowed case, we use our assumed mode number $m=2$.  These estimates agree well with the analytical calculations. Notably, $\delta v_{\phi}$ in the planetary case  is much smaller than that in the shadowed case, which is due to $\tilde{\omega}\propto m$ appearing in the denominator of $v_{\phi,1}$. On the other hand, observations are more likely to capture strong spirals in their nonlinear phase, where these relationships may no longer hold. Distinguishing between planetary spirals and shadow-induced spirals will likely prove challenging.

\subsection{Simulations with Different Disk and Shadow Parameters}
\begin{figure}[t!]
\includegraphics[trim=0mm 0mm 0mm 0mm, clip, width=3.3in]{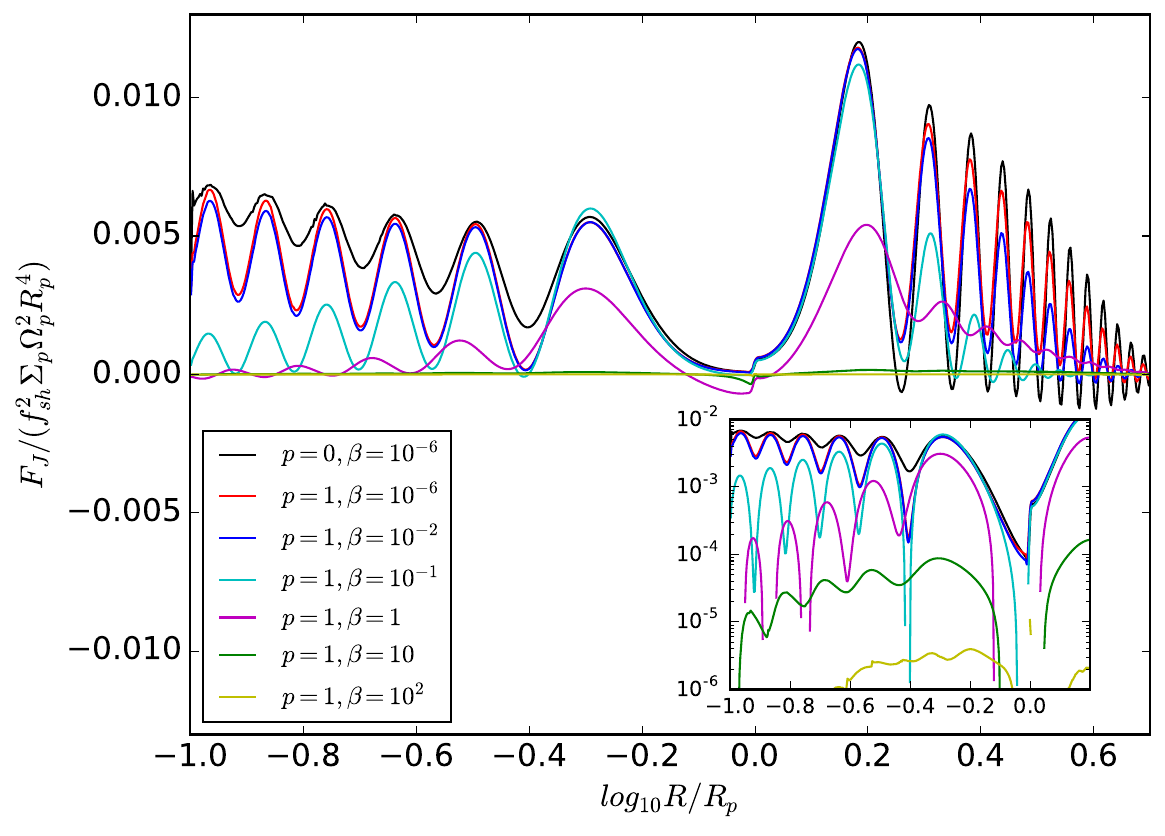}
\figcaption{Angular momentum flux in simulations with different cooling times. The m=2 shadow temperature is applied. The inset shows quantities in the logarithmic scale.
\label{fig:tctorque}}
\end{figure}

Given that our numerical simulations closely reproduce the analytical results, we will use numerical simulations in this section to explore the effects of surface density gradients, cooling times, sharpness of shadows, and amplitude of temperature perturbations. 

We first investigate how the surface density profile affects the results. As shown in Figure \ref{fig:tctorque}, the case with p=1 yields an $F_J$ similar to that of the p=0 case. Thus, the angular momentum flux is not sensitive to the slope of the surface density profile. 

Next, we examine how cooling affects $F_J$. For $\beta\lesssim10^{-2}$, the results are nearly identical to the locally isothermal case. When $\beta=0.1$, the angular momentum flux decreases toward the inner disk, being approximately five times smaller at $R/R_p=0.1$. When $\beta=1$, the angular momentum flux is roughly half of the locally isothermal value at $R\sim 0.5$ and $1.5$, but it decreases rapidly toward the inner disk, becoming nearly two orders of magnitude smaller at $R/R_p=0.1$. For $\beta\gtrsim$10, the angular momentum flux is reduced by more than two orders of magnitudes throughout the disk. The weaker spirals due to a long cooling are consistent with recent simulations by \cite{su24}. 

\begin{figure*}[t!]
\includegraphics[trim=0mm 0mm 0mm -6mm, clip, width=7in]{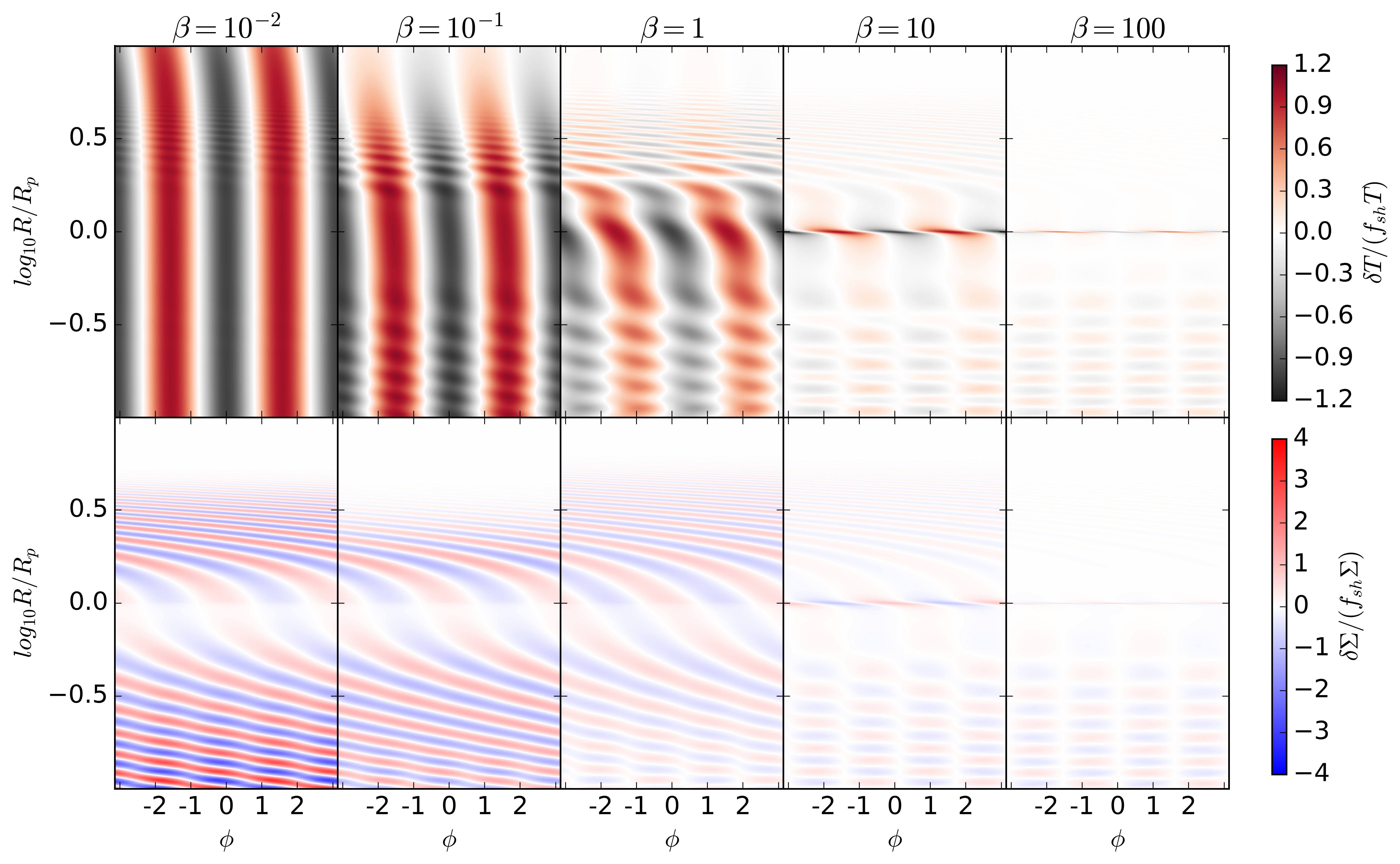}
\figcaption{Temperature (upper panels) and density (lower panels) perturbations for simulations with different cooling time (from left to right panels). 
\label{fig:tcmap}}
\end{figure*}

The two-dimensional temperature and density perturbations for simulations with different $\beta$ values are shown in Figure \ref{fig:tcmap}. In the outer disk, we observe that the temperature perturbations are affected by Keplerian motion, with slower cooling leading to a larger azimuthal shift. In the inner disk, both temperature and density perturbations exhibit a checkerboard pattern.  However, this checkerboard pattern becomes weaker spiral patterns if the temperature at our inner boundary is reset to the initial temperature without azimuthal variation. 
As will be shown in a subsequent paper (Zhu \etal, in prep), whether the disk feature follows the checkerboard pattern or spiral pattern is strongly affected by the inner boundary condition when there is little net angular momentum transport within the disk.

Azimuthal temperature variations in protoplanetary disks can exhibit different amplitudes and widths. For example, a slightly misaligned inner disk can cast a broader and more gradual shadow onto the outer disk \citep{Facchini2018}. Thus, we next adopt different shadow profiles (shown in Figure \ref{fig:sizetemperature}) in the simulations to explore their effects. We apply $\beta=10^{-6}$ and $f_{sh}=0.01$ in all these cases.

\begin{figure*}[t!]   
\includegraphics[trim=0mm 0mm 0mm -6mm, clip, width=7in]{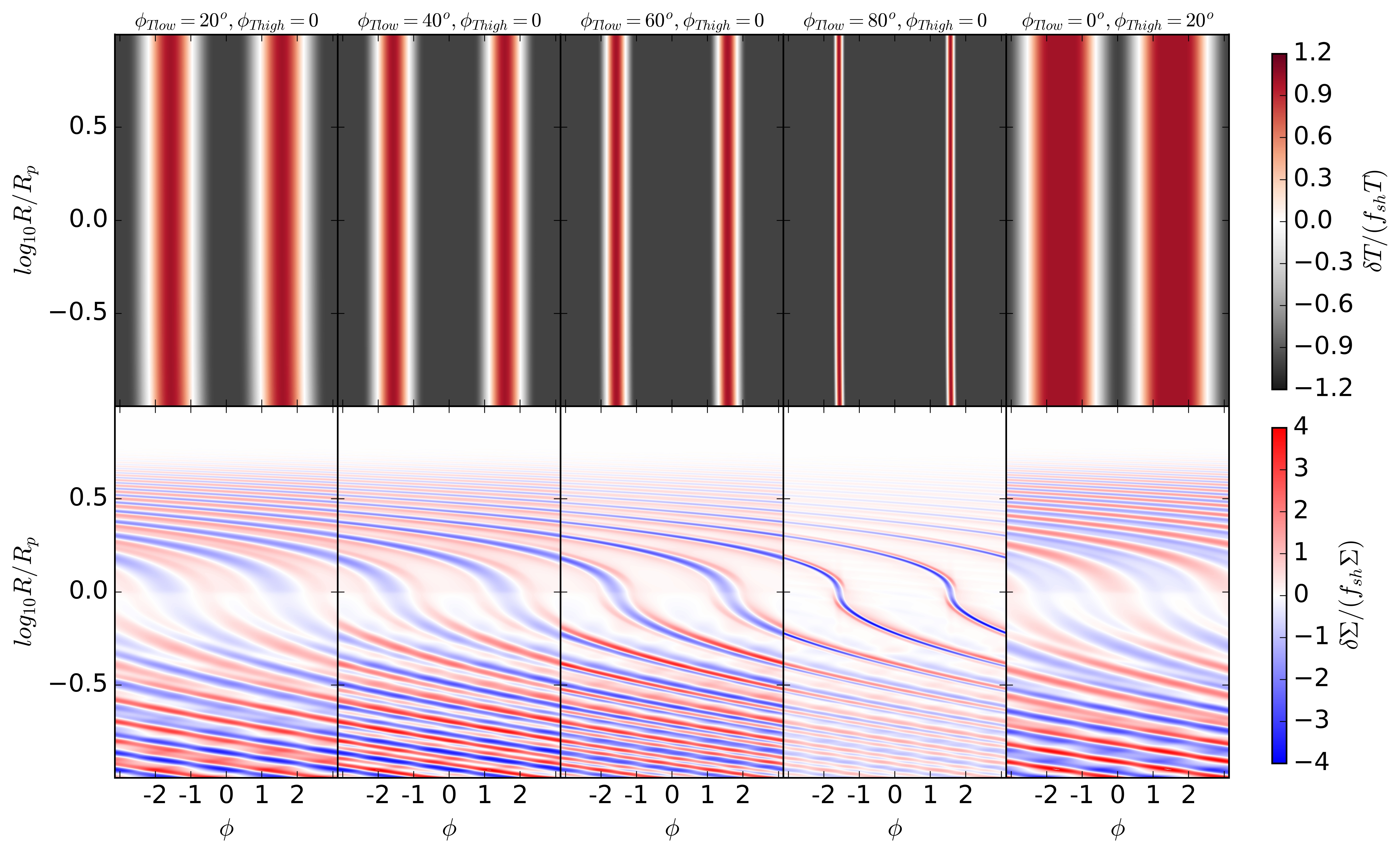}
\figcaption{Temperature (upper panels) and density (lower panels) perturbations for simulations with different azimuthal temperature profiles (from left to right panels). 
\label{fig:sizemap}}
\end{figure*}

The temperature and density perturbations are shown in Figure  \ref{fig:sizemap}. In cases where the values of $\phi_{T_{low}}$ and $\phi_{T_{high}}$ are exchanged (the leftmost and rightmost panels), the perturbation from one case can be transformed into that of the other case by flipping the sign of the perturbed values and shifting all quantities azimuthally by 90$^o$, as predicted by linear theory. As the transition becomes sharper from the left to the right panels, the spirals become narrower, and higher m modes excite stronger spirals closer to the corotation radius. Consequently, the spirals in cases with sharper transitions more closely resemble those excited by a planet. 

\begin{figure}[t!]
\includegraphics[trim=0mm 0mm 0mm 0mm, clip, width=3.3in]{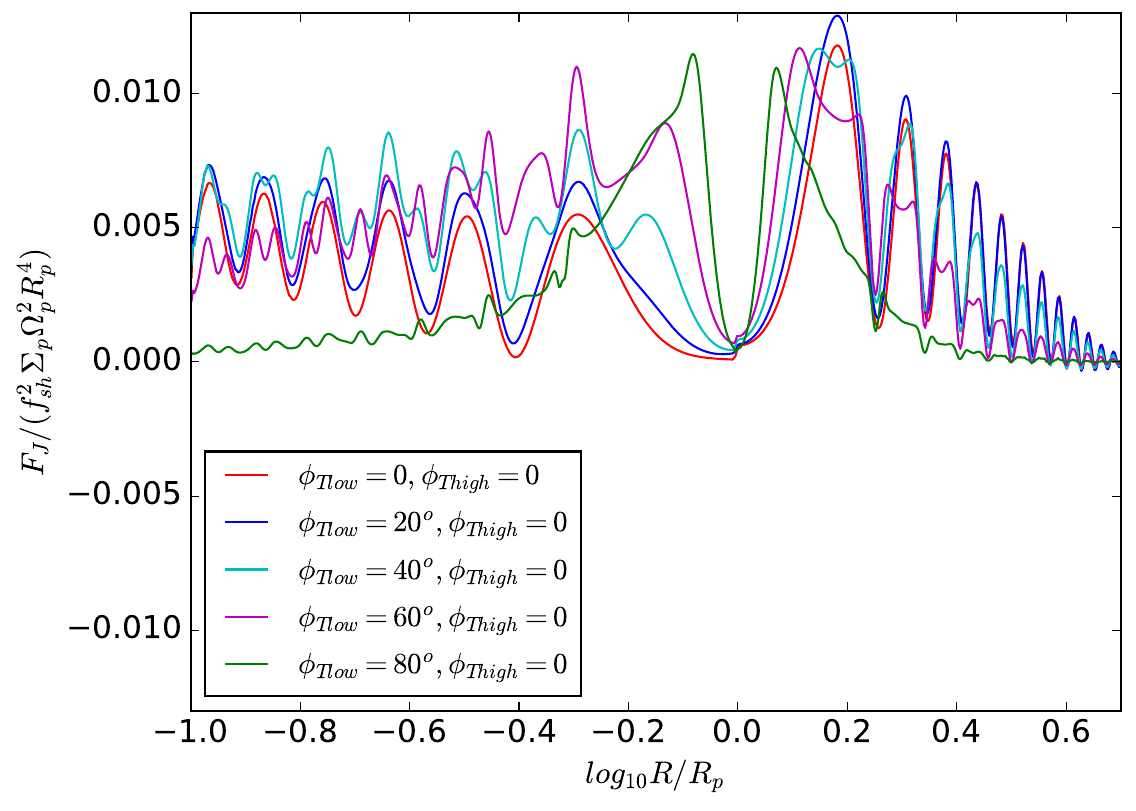}
\figcaption{Angular momentum flux in simulations with different azimuthal temperature profiles. The m=2 shadow temperature is applied. 
\label{fig:sizetorque}}
\end{figure}

\begin{figure}[t!]
\includegraphics[trim=20mm 10mm 29mm 0mm, clip, width=3.3in]{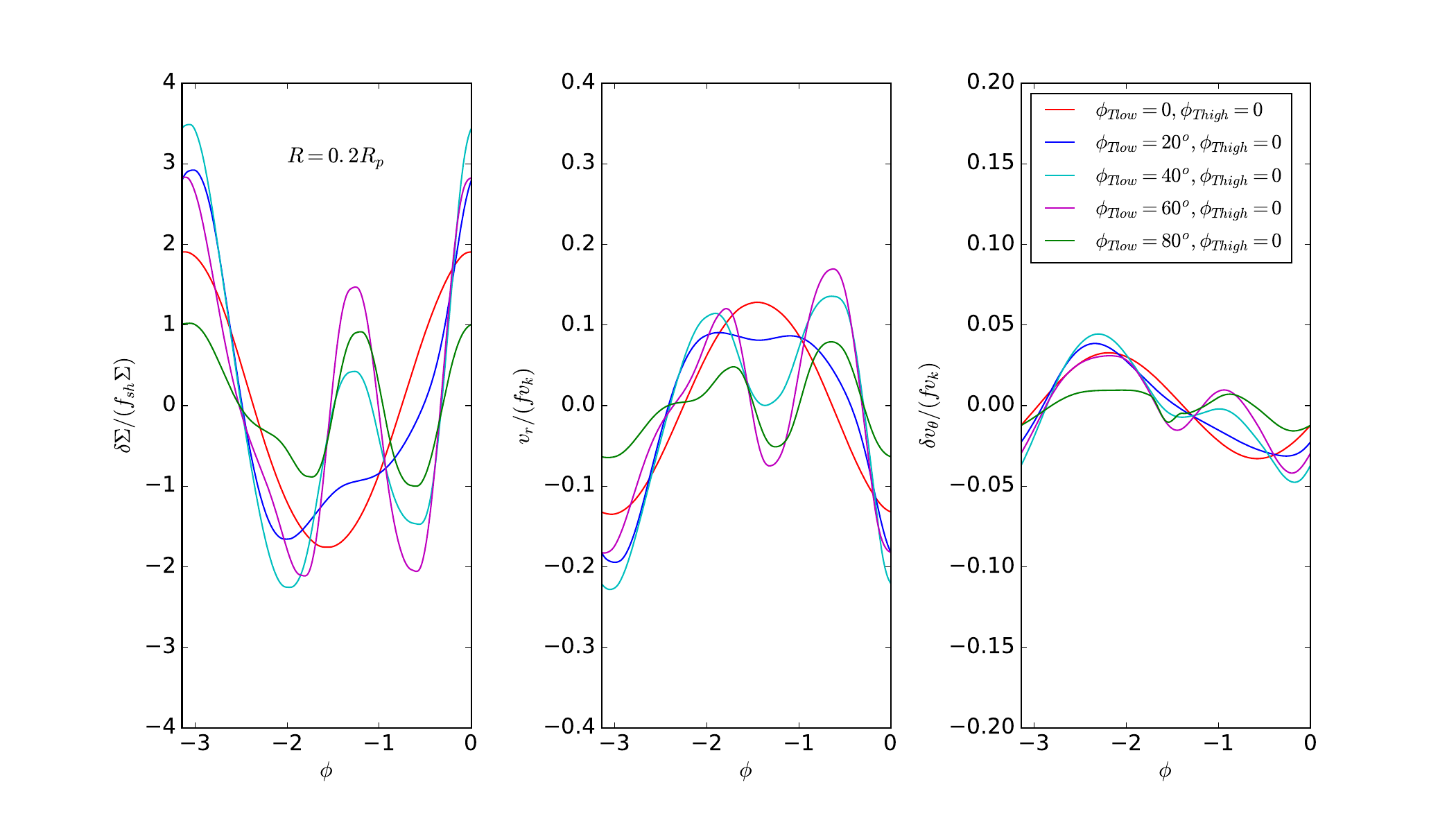}
\figcaption{The density and velocity perturbations along the $\phi$ direction at $R=0.2 R_p$ for simulations with different azimuthal temperature profiles. 
\label{fig:sizeper}}
\end{figure}

The amplitudes of the angular momentum fluxes for these cases are shown in Figure \ref{fig:sizetorque}. The sharper the shadow, the higher $m$ mode it excites, causing the first flux peaks (adjacent to $R_p$) to move closer to $R_p$. When the shadow becomes sharp, as in cases where $\phi_{T_{low}}\gtrsim60^o$, the flux decreases toward smaller radii in the inner disk. This behavior is likely due to wave steepening and shock dissipation \citep{Goodman2001}. The density and velocity perturbations along the azimuthal direction at $R=0.2 R_p$ are shown in Figure \ref{fig:sizeper}. Overall, the azimuthal profiles become sharper for sharper shadows. An additional peak besides the $m$=2 mode becomes apparent when $\phi_{T_{low}}\gtrsim40^o$. On the other hand, the amplitudes of the perturbations remain within a factor of three across cases with different shadow sharpness, indicating they are relatively insensitive to the sharpness of the temperature perturbation. The low m mode perturbations still dominate, and the amplitude seems to mostly depend on $f_{sh}$.

\begin{figure}[t!]
\includegraphics[trim=0mm 0mm 0mm 0mm, clip, width=3.3in]{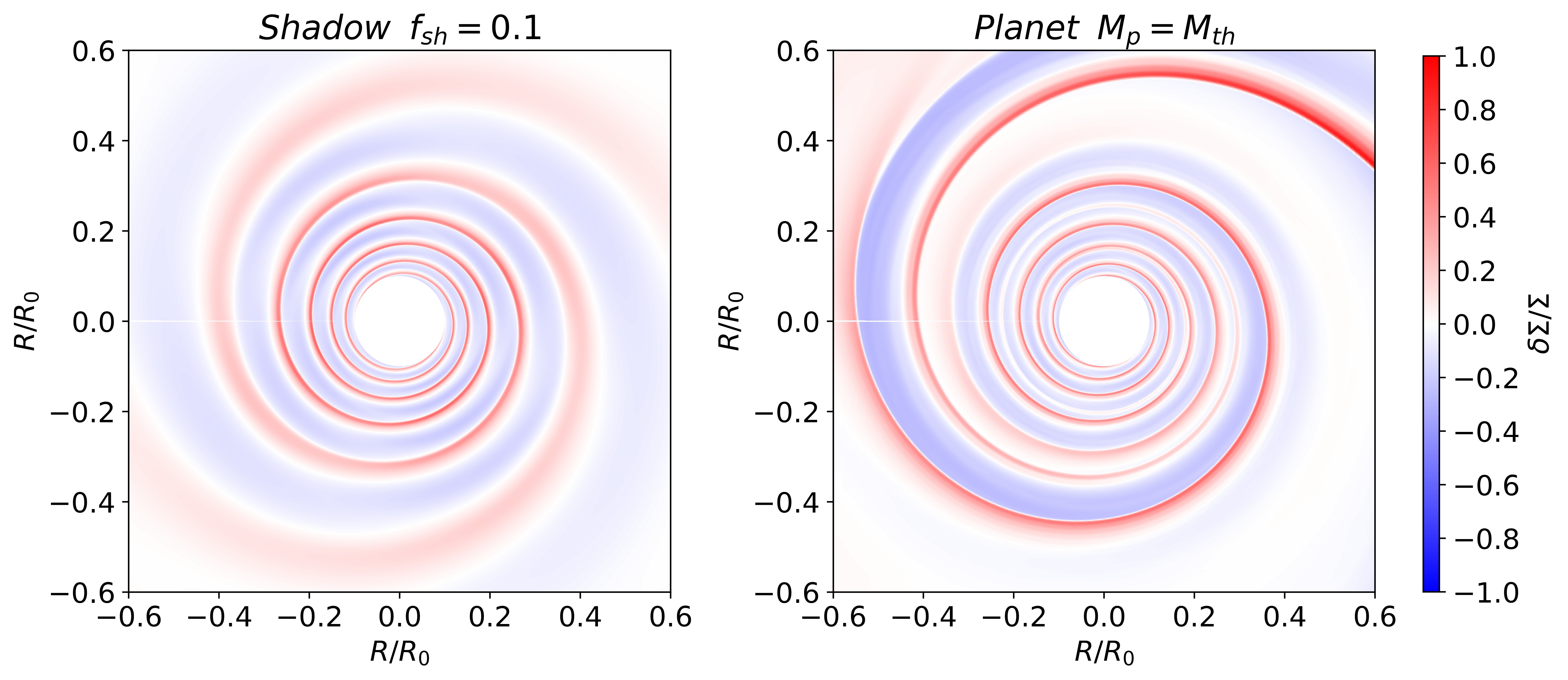}
\figcaption{The density perturbations for the $f_{sh}=0.1$ case and the $M_p=M_{th}$ case. The snapshots are taken at 2 orbital times at $R_0$ after adding the shadow or planets.
\label{fig:f0p9twod}}
\end{figure}

\begin{figure}[t!]
\includegraphics[trim=0mm 0mm 0mm 0mm, clip, width=3.3in]{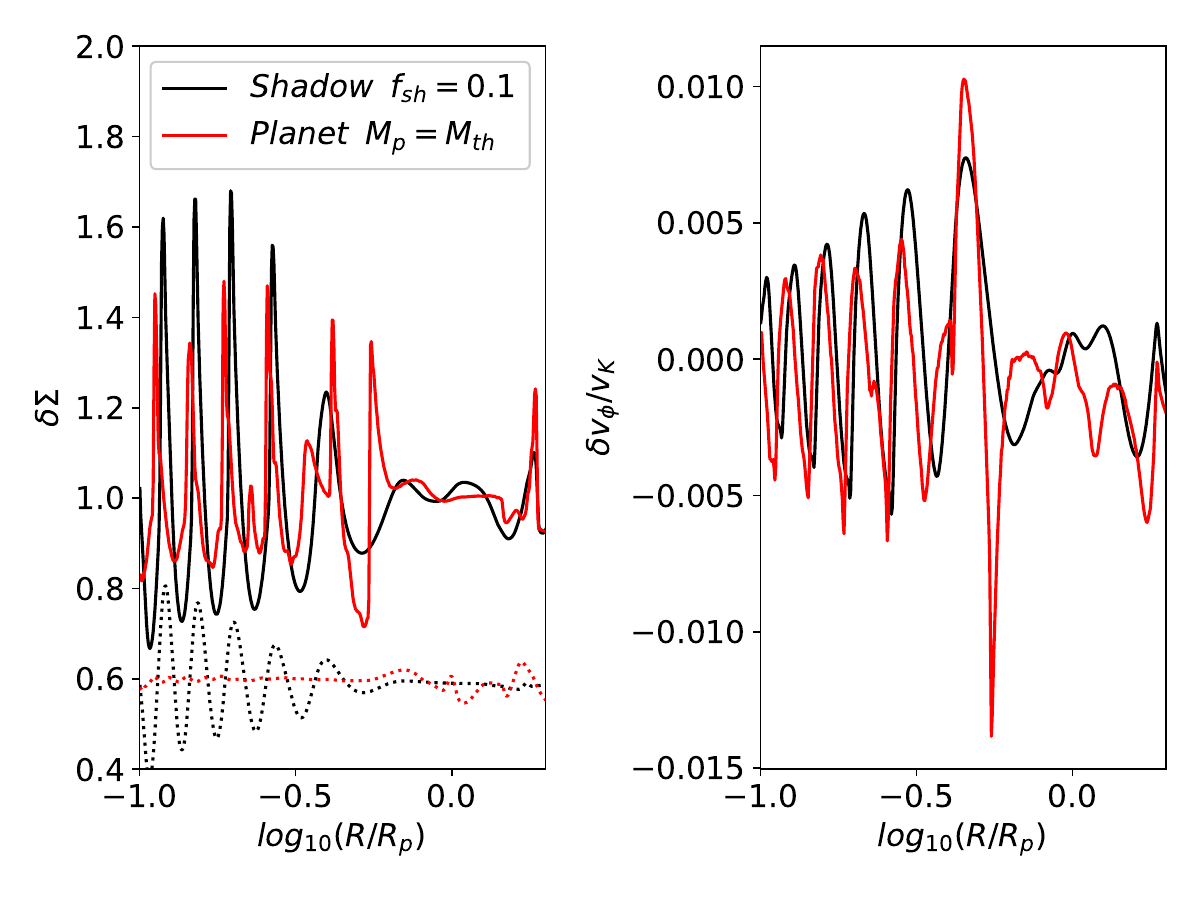}
\figcaption{Solid curves: the density and $v_{\phi}$ perturbations for the $f_{sh}=0.1$ case and the $M_p=M_{th}$ case along $\phi=0$. The dotted curves in the left panel are the azimuthally averaged density perturbations. The curves are shifted by -0.4 for clarity. All data are from Figure \ref{fig:f0p9twod}.
\label{fig:f0p9d}}
\end{figure}

\begin{figure}[t!]
\includegraphics[trim=0mm 0mm 0mm 0mm, clip, width=3.3in]{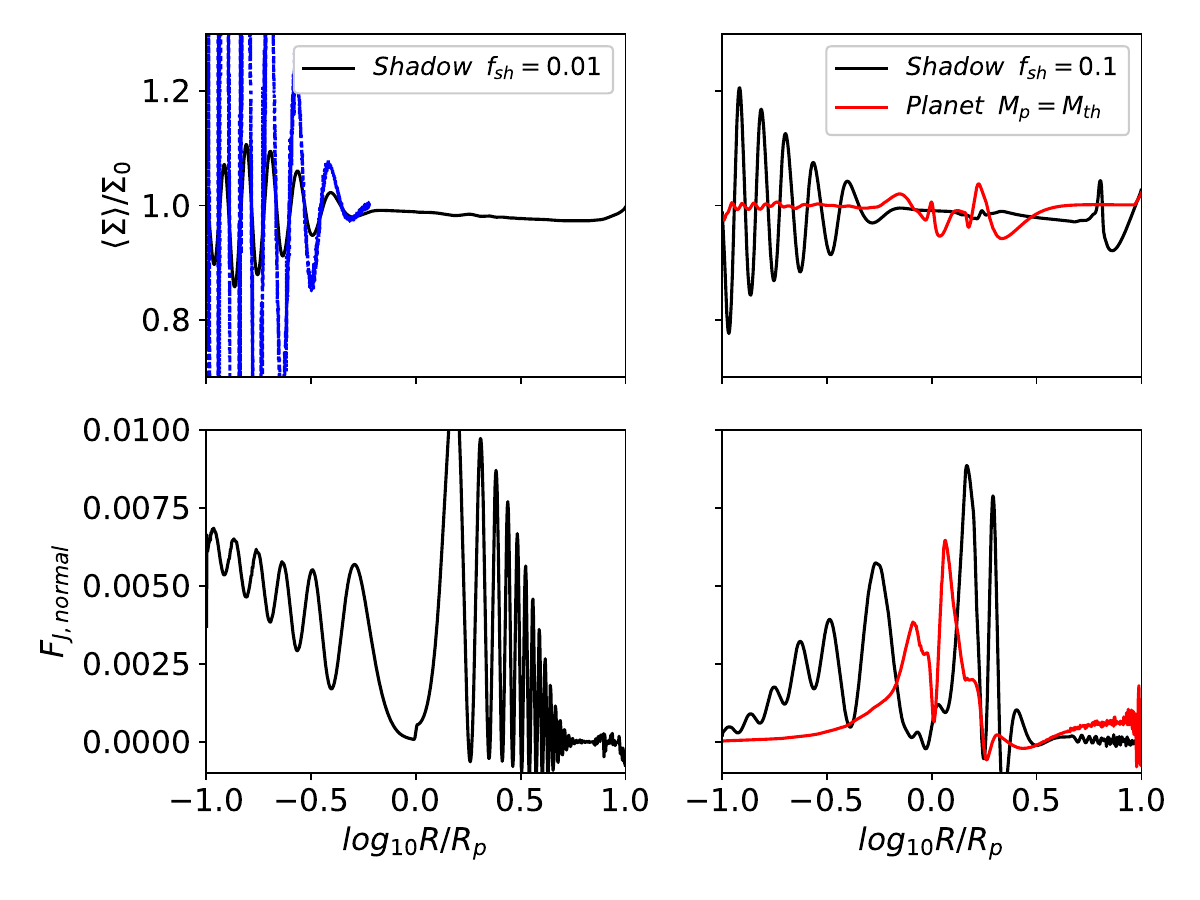}
\figcaption{The azimuthally averaged density (upper row) and angular momentum flux (bottom row) for weak shadows at 100 orbits (left column) and strong shadows at 2 orbits (right column). Using Equation \ref{eq:Sigmaevo} and $F_J$ from the linear analysis, we calculate the surface density at 100 orbits, shown as the blue curve in the upper left panel.
The thermal mass planet case is also shown in the right panels (red curves). The angular momentum flux is normalized in the same way as that in Figure \ref{fig:torque}.
\label{fig:avgsigma}}
\end{figure}

In a realistic protoplanetary disk that has shadows cast by the inner disk, the temperature perturbation can reach order of unity. 
To study the response of the disk to a stronger azimuthal temperature variation, we increase $f_{sh}$ from 0.01 to 0.1. Based on the profiles shown in Figure \ref{fig:linearcuts} from the linear theory, the density perturbation will reach unity when $f_{sh}=0.1$. The resulting density perturbation at 2 orbits (at $R_p$) after adding the shadow is shown in the left panel of Figure \ref{fig:f0p9twod}, which confirms that even $f_{sh}=0.1$ is sufficient to drive spiral structures of order unity. For comparison, we have also run a hydrodynamical simulation with a $M_{th}$ planet in the disk, and the density perturbation is shown in the right panel of Figure \ref{fig:f0p9twod}. The primary and secondary spirals \citep{Zhu2015c,Bae2018,Miranda2019} are apparent in the planetary case. The radial profiles of density and velocity perturbations for both cases are shown in Figure \ref{fig:f0p9d}. Please note that the azimuthally averaged density profiles show ring structures for the shadowed case, which will be discussed in \S 4.4.  Overall, the spirals in these two cases are quite similar in both morphology and amplitude. 

When we run the simulations for longer (especially the $f_{sh}=0.1$ case), we notice that the rings in the shadowed case quickly become unstable and generate oscillatory spirals. We suspect that the density rings are subject to the Rossby Wave Instability \citep{lovelace99} and generate vortices. The resulting spirals from vortices could have pattern speeds different from $\Omega_p$. More spirals and gaps are also evident. 

\subsection{Ring Formation and Disk Accretion}
After studying the angular momentum flux in the disk, we are ready to understand the ring formation.
The change of the angular momentum flux leads to disk accretion. We can integrate the angular momentum equation  along the azimuthal direction,
\begin{equation}
\frac{\partial R\Sigma v_{\phi}}{\partial t}+\nabla_c\cdot\left(R\Sigma {\bf v}v_{\phi}+P{\bf e_{\phi}}\right)=\frac{dT}{dR}\,,
\end{equation}
where $\nabla_c$ is the divergence operator in cylindrical or spherical polar coordinates, and the torque density from any external potential is
\begin{equation}
\frac{dT}{dR}=-R\int\Sigma\frac{\partial \Phi_{ext,1}}{\partial \phi}d\phi\,.
\end{equation}
With the mass conservation equation, we have
\small
\begin{equation}
\frac{\partial \int R\Sigma \delta v_{\phi}d \phi}{\partial t}=-\frac{1}{R}\frac{\partial}{\partial R}\left(R^2\int\Sigma v_{R}\delta v_{\phi}d\phi -\frac{dT}{dR}\right)-\frac{\dot{M}}{ R}\frac{\partial R v_{k}}{\partial R}\,,
\end{equation}
\normalsize
where $\delta v_{\phi}=v_{\phi}-v_k$, and $\dot{M}=\int \Sigma v_{R} R d\phi$. Thus,
\begin{equation}
\dot{M}=-\frac{2}{v_{k}}\left(\frac{d F_J}{d R}-\frac{dT}{dR}\right)-\frac{2R^2}{v_k}\frac{\partial \int\Sigma \delta v_{\phi}d \phi}{\partial t}\,.\label{eq:mdotevo}
\end{equation}
In our cases with no external potential and the only driving force is the azimuthal temperature variation, we have
\begin{equation}
\dot{M}=-\frac{2}{v_{k}}\frac{d F_J}{d R}-\frac{2R^2}{v_k}\frac{\partial \int\Sigma \delta v_{\phi}d \phi}{\partial t}\,.\label{eq:mdotevo}
\end{equation}
 For a steady state, the second term on the right-hand side becomes zero. However, a steady state is not possible in the shadowed disk case where $F_J$ is oscillatory. Assuming it achieves a steady state, the radially oscillatory $F_J$ would lead to a radially oscillatory $\dot{M}$ which would build up rings and gaps, violating the steady state assumption. Thus, structure has to emerge in the disk. Figures \ref{fig:f0p9d} and \ref{fig:avgsigma} indeed show that the radial profile of the azimuthally averaged surface density have rings that correspond to the variations of $F_J$.
 
This accretion due to the wave's angular momentum flux change is called the ``anomalous'' wave-driven accretion by \cite{Miranda2020}. With $\dot{M}$ known, the disk's surface density evolution can then be derived with
\begin{equation}
\frac{\partial \Sigma}{\partial t}=-\frac{1}{2\pi R}\frac{\partial{\dot{M}}}{\partial R}\,.\label{eq:Sigmaevo}
\end{equation}
Unfortunately, we need to know the second order terms in $\Sigma\delta v_{\phi}$ (Equation \ref{eq:mdotevo}) for calculating $\dot{M}$, which is beyond our linear analysis. Nevertheless, we can assume that the time derivative term in Equation \ref{eq:mdotevo} is zero, and solve Equations \ref{eq:mdotevo} and \ref{eq:Sigmaevo} using $F_J$ that has been derived analytically. This exercise leads to the blue curve in Figure \ref{fig:torque}. The blue curve is only shown for the inner disk to avoid the singularity at the corotation radius. The blue curve is roughly twice the surface density from the simulation (the black curve), suggesting that the time derivative term in Equation \ref{eq:mdotevo} (which is also second order, the same as the angular momentum flux term) is not negligible. Nevertheless, using only the wave's angular momentum flux produces the density peaks at the correct radii. We leave a more accurate solution with second order terms to future works.

This ring formation mechanism is dramatically different from the ring/gap formation mechanism in planet-disk interaction. Shown as the red dotted curve in Figure \ref{fig:f0p9d} and red solid curves in Figure \ref{fig:avgsigma},  the planetary case does not produce significant rings in the same amount of time.  In an adiabatic disk, the planet's gap-opening process has three steps: wave excitation by the planetary torque, wave propagation, and wave nonlinear steepening/dissipation. Starting from the Lindblad resonances, the planet torques the spiral waves and  injects angular momentum to the waves. When the waves leave Lindblad resonances, the coupling between the waves and the planetary potential is weak mainly because the planetary potential decreases sharply with the distance from the planet. Then, the spiral waves propagate freely and conserve angular momentum flux in adiabatic disks \citep{Toomre1969, GoldreichTremaine1979}. During the propagation, the spiral waves will gradually steepen to weak shocks, and dissipate in the disk \citep{Goodman2001}. Angular momentum of the wave is then transferred to the disk, leading to gap opening (Equation \ref{eq:mdotevo}). This gap opening process is slightly different for a planet in a locally isothermal disk. After the wave excitation, the free-propagating waves do not conserve angular momentum in a locally isothermal disk. Instead they conserve $F_J/c_s^2$ \citep{Lin2011,Miranda2020}. Since the total angular momentum of the disk has to be conserved, the change of $F_J$ leads to disk accretion (Equation \ref{eq:mdotevo}). This accretion process does not require a non-linear wave in the disk, and only requires a free-propagating wave. Thus it is called anomalous wave-driven accretion. In reality, the waves excited by an earth mass planet are strong enough that  non-linear wave steepening will occur quite quickly after the waves are launched, and this anomalous wave-driven accretion is not apparent. 

In our case with azimuthal temperature perturbations, there is no external torque since pressure is an internal force. Thus, angular momentum flux is conserved from the start. While the azimuthal temperature perturbations drive spirals, the background disk has to transport mass through the anomalous wave-driven accretion (Equation \ref{eq:mdotevo}). Since the temperature perturbations exist throughout all the disk, they couple with the waves at all radii. There is no free-propagating wave, and the angular momentum flux of the waves oscillates throughout the disk. This oscillation drives ring formation. Finally, we note that the ring formation due to the oscillatory $F_J$ may also apply to circumprimary disks in binary systems or circumplanetary disks in planet-star systems, where the companion's gravitational potential remains strong far away from the companion to couple with low-m spirals excited by the companion. The rings observed in circumplanetary disks (e.g., Figure 5 in \citealt{Zhu2016} and Figure 4 in \citealt{Chen2021}) may be attributed to this mechanism too.

One may ask why the spirals produced by shadows have not steepened into shocks and dissipated in the disk. This is mainly because we start with a low amplitude m=2 perturbation. As shown in \S 4.3, a sharper and deeper shadow, which is more realistic, could lead to radially decreasing  angular momentum flux, likely due to non-linear wave steepening and dissipation. \cite{Goodman2001} have shown that spirals launched by embedded planets may lead to moderate accretion through this non-linear wave dissipation. They have used the shock dissipation length to estimate $dF_J/dR$. Although the dissipation length in the shadowed case depends on the shadow's sharpness, we assume that it is on the order of $R_p$ as a rough estimate.  Based on Equation \ref{eq:fshadow}, we could estimate $\dot{M}\sim 40 f_{sh,p}^2\Sigma_p c_{s,p}^4 \Omega_p^{-3} R_p^{-2}$. If we consider $\dot{M}=3\pi R\Sigma \alpha c_{s}^2/(\Omega R)$, we have
\begin{equation}
\alpha\sim 5 f_{sh}^2 h^2\,.
\end{equation} 
Thus, if $f_{sh}\sim 0.1$ and $h\sim 0.1$, $\alpha\sim 10^{-3}-10^{-4}$. On the other hand, our estimate may be too simplified. If the wave is dissipated at a radius that is much smaller than $R_p$, the equivalent $\alpha$ can be very different. For example, if the wave is dissipated in the cavity of the transitional disk with a small $\Sigma$ (as in \citealt{Zhang2024b}), $\alpha$ can be a lot higher. Overall, both anomalous wave-driven accretion and wave steepening dissipation-driven accretion may occur simultaneously in the shadowed disks. The former likely leads to rings and the latter leads to overall accretion.

\section{Discussions}
\subsection{Cooling timescale}
Our results indicate that slow cooling with $\beta>1$ significantly suppresses  wave excitation (also in \citealt{su24}). This phenomenon can be understood as a competition between advection and  cooling processes. The timescale for a Keplerian fluid element to traverse one shadow is given by
\begin{equation}
t_{cross}=\frac{2\pi}{m|\Omega_p-\Omega|}\,.
\end{equation}
During this time, the irradiation temperature decreases, allowing the disk to cool. If the disk's cooling time is shorter than this crossing time, 
\begin{equation}
t_{cool}<\frac{2\pi}{m|\Omega_p-\Omega|}\,,
\end{equation}
the disk responds to the drop in irradiation temperature and spiral waves are launched. When $R<<R_p$ and $m\sim 2$,
this condition simplifies to $t_{cool}\Omega<\pi$ or $\beta<\pi$. Therefore, the efficiency of wave excitation by shadows depends on the cooling time -- a contrast to wave excitation by a planet, where the planetary potential excites waves with similar $F_J$ independent from the cooling time.

Although our two-dimensional simulations cannot study the disk's vertical motion, we expect that slow cooling could also dampen the vertical motion of disk fluid within the shadow.
If the cooling time is shorter than the orbital time, the disk loses thermal support within the shadow. However, a new hydrostatic equilibrium can not establish instantaneously because both the thermal timescale ($H/c_{s}$) and the free-fall timescale are on the order of the orbital timescale. As a result, the disk oscillates vertically \citep{Benisty2017,Zhang2024b}, which may have observational implications. On the other hand,
if the cooling time is too long, the disk remains adiabatic while crossing the shadow, maintaining hydrostatic equilibrium with little vertical motion. Thus, slow cooling will prevent the shadows from generating any disk features in all three dimensions. 

In realistic protoplanetary disks, the cooling time depends on factors such as disk opacity, optical depth, and temperature (e.g. Equation 9 in \citealt{Zhu2015c}). Typically, $\beta\lesssim 1$ is satisfied at distances of tens of au. Thus, shadow-induced spirals are expected to occur in outer disks, and may explain the spirals observed in some large transitional disks.

\subsection{Observational Implications}
In this work, we  present a linear theory to explain previous simulations that an orbiting temperature perturbation could excite spirals and rings \citep{montesinos18,su24}. We  show that temperature perturbations are efficient in exciting spirals in disks. Even a 10$\%$ temperature variation can excite spirals with an order of unity density change. Since these are still density waves in disks, the temperature-driven spirals have the same pitch angle dependence as the spirals excited by planets.
However, temperature-driven spirals differ from planet-driven spirals in several ways. 

Firstly, for weak spirals that are in the linear phase, the number and structure of the spirals are directly linked to the specific profiles of the temperature variations. Given that the temperature variations are typically dominated by low-m modes (e.g. two shadows cast by an inner disk), temperature-driven spirals tend to be broader in both radial and azimuthal extends compared to planet-driven spirals which have dominant modes around $m\sim1/h$. These temperature-driven spirals also exhibit much stronger azimuthal velocity perturbations (Figure \ref{fig:linearcuts}).  However, for stronger spirals, where density perturbation approach unity, the spiral structure is shaped by nonlinear wave-steepening and shock dissipation (e.g. \citealt{Goodman2001,Bollati2021}). In this regime, temperature- and planet-driven spirals appear more similar (Figure \ref{fig:f0p9d}). 

Secondly, since temperature variations (e.g., due to shadows) often extend beyond a single corotation radius, the coupling between temperature perturbations and spirals occurs throughout the disk, leading to radially oscillatory angular momentum flux. This promotes ring formation throughout the disk (also true for stationary shadows, Zhu et al. in prep.), which can further lead to vortex formation and disrupt the spiral structures. 

Finally, the pattern speed of the spirals is related to the orbital frequency of the temperature variations, which are likely to vary with time. If the temperature variation is due to the shadow cast by a misaligned inner disk, the precession of this inner disk can introduce a time-varying perturbation. The movement of the shadow that is cast by the inner disk onto the outer disk could be quite complex, depending on the precession axis and the relative inclination between the inner and outer disks (equation 54 in \citealt{Zhu2019}). Depending on these factors, the shadow may sweep across the outer disk in a circular motion or oscillate around fixed position angles. Nevertheless, the precession frequency can still provide a rough estimate of the orbital frequency of the shadow's movement. 

Here we consider two scenarios that drive the inner disk precession. The first scenario is that a misaligned inner disk is separated from an outer disk by a low mass companion (e.g. a planet) at $R_p$  with the mass ratio $q\equiv M_p/M_*$. The companion exerts a torque to the inner disk. The precession frequency (Equation 27 in \citealt{Zhu2019}) is 
\begin{equation}
\frac{\omega_{s}}{\Omega_p}\sim -\frac{7.5-3p}{16-4p}q\left(\frac{1}{1+q}\right)^{1/2}\left(\frac{R_{out}}{R_p}\right)^{3/2}\,,
\end{equation}
where the circumprimary disk is from 0 to $R_{out}$ with the surface density profile $\Sigma(R)\propto R^{-p}$. Assuming $p=1$, and $R_{out}\sim 0.5 R_p$, we have
\begin{equation}
\frac{\omega_{s}}{\Omega_p}\sim 0.1 \frac{M_{p}}{M_*}\,.
\end{equation}
Thus, if the inner disk's shadow has the corotation radius of 100 au or $\omega_s=\Omega_{k}(100 au)$, the companion could be a Jupiter mass planet at 0.2 au, or a 10 Jupiter mass planet at 1 au. On the other hand, the inner and outer disks need to be significantly misaligned or anti-aligned for the spirals moving in the same direction as the outer disk's rotation. 
If the inner and outer disks are mildly misaligned, the inner disk precesses counter-clockwise with respect to the outer disk's rotation, and the precession does not have Lindblad resonances in the outer disk.

The second scenario is that a misaligned inner circumbinary disk precesses around the binary angular momentum vector and the precessing inner disk casts shadows to the outer disk.
The precession frequency is
\citep{Lubow2018}
\begin{equation}
\frac{\omega_{s}}{\Omega_b}=\frac{|k|(5-2p)a_{b}^{7/2}}{2(1+p)r_{in}^{1+p}r_{out}^{5/2-p}}
\end{equation}
where $k$ is a constant (Appendix A in \citealt{Lubow2018}), $a_b$ is the semi-major axis of the binary, and the inner circumbinary disk is from $r_{in}$ to $r_{out}$ with the surface density profile $\Sigma(R)\propto R^{-p}$. If we assume $p=1$, $r_{in}\sim 2a_b$, and the equal mass binary, we have
\begin{equation}
\omega_{s}=0.1\Omega_{out}\,,
\end{equation}
where $\Omega_{out}$ is the orbital frequency at the disk outer radius $r_{out}$. If we again assume that the inner disk's shadow has the corotation radius of 100 au, the inner disk should then extend to $\sim$20 au around the equal mass binary. Similarly, the inner and outer disks need to be significantly misaligned so that the precession is in the same direction as the outer disk's rotation. 

In reality, the shadow's corotation radius may not be within the protoplanetary disk (e.g. a stationary, a very slowly precessing, or a retrograde shadow) and steady spirals may not be launched at Lindblad resonances. However, as will be shown in the subsequent paper (Zhu \etal in prep),  spirals can still be launched although they depend on the boundary conditions applied. Many of our orders of magnitude estimates here can still be applied.

\section{Conclusion}
Protoplanetary disks can exhibit asymmetric temperature variations due to phenomena such as shadows cast by the inner disk or localized heating by young planets. We have performed linear analyses to investigate the disk's response to the asymmetric temperature variations, particularly on spiral and ring formation. Spiral amplitudes are proportional to the amplitude of the temperature variation ($f_{sh}$). In the special case that the radial density gradient is zero, the temperature perturbation acts through an equivalent potential, analogous to the gravitational potential of a perturbing body.  

The linear analysis and direct numerical simulations show excellent agreement and demonstrate that the effects of temperature variations share similarities with those caused by external gravitational potentials. Specifically, rotating temperature variations launch steady spiral structures at Lindblad resonances, which corotate with the temperature patterns. The only significant difference is that, since the temperature perturbation is non-zero at all radii, it can couple with the spiral patterns and lead to radially varying angular momentum flux throughout the disk. Nevertheless, using the smoothed component of the angular momentum flux, we provide expressions for the amplitudes of the resulting density and velocity perturbations. Compared to planet-driven spirals, temperature-driven spirals tend to be broader in both radial and azimuthal extents and also exhibit much stronger azimuthal velocity perturbations due to its low-m modes. 

When the cooling time exceeds the orbital period, these spiral structures are significantly weakened and the angular momentum transport is diminishing. Depending on the boundary conditions, the disk can exhibit a checkerboard pattern instead. 

While the azimuthal profiles of the density perturbations are influenced by the azimuthal profiles of the temperature variations, the amplitude of the density perturbations is primarily determined by the magnitude of the temperature variations. Notably, a temperature variation of about 10\% can induce spirals with amplitudes of order unity, comparable to those generated by a thermal mass planet. 

The radially varying angular momentum flux that is from the coupling between temperature variations and spirals outside the resonances could result in other disk substructures (e.g., rings, vortices)  due to wave-driven accretion. This ring formation mechanism is different from the non-linear wave steepening/dissipation in planetary gap opening. On the other hand, we speculate that spirals induced by large amplitude temperature variations may also lead to disk's overall inward accretion through non-linear wave steepening and dissipation. Both anomalous wave-driven accretion and wave steepening dissipation-driven accretion may occur simultaneously in the shadowed disks. The former likely leads to rings and the latter leads to overall accretion. 

Overall, the change of irradiation both spatially or/and temporarily may leave observable effects (spirals and rings) on protoplanetary disks, especially at the outer disk beyond tens of au where cooling is fast. 

\acknowledgments

Z. Z. thank Ryan Miranda for making his linear analysis code publicly available. We thank the anonymous reviewer for helpful comments. 
Some simulations are carried out using with NASA Pleiades supercomputer. Z. Z. acknowledges support from NASA award 80NSSC22K1413 and NSF award 2408207. S.Z. acknowledge support by NASA through the NASA Hubble Fellowship grant \#HST-HF2-51568 awarded by the Space Telescope Science Institute, which is operated by the
Association of Universities for Research in Astronomy, Inc., for NASA, under contract NAS5-26555.

\end{document}